\documentclass[10pt]{revtex4}
\usepackage{amssymb}
\usepackage{latexsym}
\usepackage{epsfig}
\usepackage{float}
\usepackage{graphicx,epsfig, color }
\usepackage{graphicx}
\usepackage{subfigure}
\usepackage{amsmath}

\begin{document}
\title{Physical and thermodynamic properties of quartic quasitopological black holes and rotating black branes with nonlinear source}

\author{ A. Bazrafshan$^{1}$\footnote{Corresponding author}, F. Naeimipour$^{2}$, M. Ghanaatian$^{2}$, A. Khajeh$^{2}$,}
\address{$^1$ Department of Physics, Jahrom University, 74137-66171 Jahrom, Iran\\
$^2$ Department of Physics, Payame Noor University (PNU), P.O. Box 19395-3697 Tehran, Iran}

\begin{abstract}
In this paper, we find the solutions of quartic quasitopological black holes and branes coupled to logarithmic and exponential forms of nonlinear electrodynamics. These solutions have an essential singularity at $r=0$. Depending on the value of charge parameter $q$, we have an extreme black hole/brane, a black hole/brane with two horizons or a naked singularity. For small values of parameter $q$, the solutions lead to a black hole/brane with two horizons. The values of the horizons are independent of the values of quasitopological parameters and depend only on the values of $q$, dimensions $n$, nonlinear parameter $\beta$ and mass parameter. Also, the solutions are not thermally stable for dS and flat spacetimes. However, AdS solutions are stable for $r_{+}>{r_{+}}_{\rm{ext}}$ which the temperature is zero for $r_{+}={r_{+}}_{\rm{ext}}$. The value of ${r_{+}}_{\rm{ext}}$ also depends on the values of parameters $q$, $\beta$, $n$ and $m$. As the value of ${r_{+}}_{\rm{ext}}$ decreases, the region of stability becomes larger. We also use HPEM metric to probe GTD formalism for our solutions. This metric is successful to predict the divergences of the scalar curvature exactly at the phase transition points. For large values of parameter $\Xi$, the black hole/brane has a transition to a stable state and stays stable.
\end{abstract}



\maketitle
\section{Introduction}
In general relativity, richer structures of higher dimensional black holes could have been attractive more than
the four dimensional black holes \cite{Myers1}. This originates from some motivations. The first is relevant to AdS/CFT
correspondence in which the dynamics of a $d$-dimensional
black hole are related to those of a quantum field
theory in $(d-1)$ dimensions \cite{Aharony}. String theory is the second motivation which is just formulated in ten dimensions. Einstein theory is merely a low energy limit
of string theory. In the low limit of energy, this theory gives rise to effective models of gravity in higher dimensions which involve higher curvature terms \cite{Boulware, Takahas}. Thirdly, stability of higher dimensional black holes is so important, since these black holes can be produced at the LHC if we consider the spacetime with dimensions larger than six \cite{Takahas}. Fourthly, as mathematical objects, black hole
spacetimes are among the most important Lorentzian Ricci-flat
manifolds in any dimensions \cite{Emparan}.\\
Einstein-Hilbert action is successful to describe the spacetime geometry in three and four dimensions, while for higher dimensions, Einstein's equations are not the most complete ones that can satisfy Einstein's assumptions. So, to extend the gravitational theories into those with higher power of curvature, we should go to the modified theories which have some corrections to the Einsteins's Lagrangian.\\
Quasitopological gravity is one of these theories which can be described in higher dimensions. This gravity has the ability to provide a useful toy model for the holographic study of four- and higher-dimensional CFT’s \cite{Myers}.
Also, in quasitopological gravity, we can find a lower non-zero value in a particular
corner of the allowed space of gravitational couplings for the ratio of the shear viscosity to entropy \cite{Myers2}. From the point of view of AdS/CFT, this gravity can also produce enough free coupling parameters to make a one-to-one relationship between central charges and couplings on the
non-gravitational side and the coupling parameters on the gravitational side \cite{Lemos1,Deh1,Mann1,Brenna,Deh2,Cai,Deh3}. Also, as the terms of quasitopological gravity are not true topological invariants, they can also produce coupling terms and nontrivial gravitational effects in fewer dimensions than other modified gravities such as Love-Lock gravity. Also, by choosing some special constraints on the coupling constants of this gravity, one can set causality for CFT \cite{Camanho,Hofman,Ge}. So, these reasons persuade us to consider quasitopological gravity in the present paper. Recently, two studies of quasitopological and quartic quasitopological gravities have been done respectively, in \cite{Myers} and \cite{Ghana2}. We have also investigated the solution of the charged black hole in quartic quasitopological gravity in \cite{Naei1}. The solutions of lifshitz quartic quasitopological black holes have been also studied in \cite{Ghana1}.\\‏
Considering nonlinear terms of invariants constructed by Riemann tensor on the
gravity side of the action, we can also add these terms to the matter part of the action, too. Nonlinear electrodynamics was introduced by the desire of removing the infinite self-energy of a point-like charge and finding non-singular field theories \cite{Born}. The other motivation of considering nonlinear electrodynamic term returns to the fact that, most of physical systems in nature, with field equations of
the gravitational systems, are intrinsically nonlinear. Recently, obtaining the solutions of quasitopological gravity in the presence of nonlinear electrodynamics has been an interesting subject to study. For example, Born-Infeld theory in the presence of quartic quasitopological gravity has been studied in \cite{Ghanna3}. We have also constructed the solutions of the cubic quasitopological black hole in the presence of power-Maxwell theory in \cite{Naeimi2}. Now, in the present paper, we tend to extend our study to the other types of nonlinear electrodynamics, namely, exponential and logarithmic forms. So, we have a purpose to construct a new class of $n$-dimensional black brane solutions in quartic quasitopological gravity coupled to nonlinear electrodynamics such as exponential and logarithmic forms.\\
The outline of our paper is organized as follows: In sec. \ref{Field}, we first introduce the nonlinear electrodynamics Lagrangians such as exponential and logarithmic ones and then define a $(n+1)$-dimensional action in quasitopological gravity with nonlinear electrodynamics. In sec. \ref{static}, we use the metric of the static spacetime and obtain the related equations and then solve them to find the static solutions, analytically. In Sec. \ref{rotating}, we extend this spacetime to a rotating case and obtain the solutions of rotating black brane. We probe the physical and thermodynamic structure of the relevant solutions, respectively, in sec. \ref{structure} and \ref{Thermo}. Sections \ref{stability} and\ref{geom} are devoted to study thermal stability and geometrothermodynamics on the obtained solutions. Finally, we present a brief conclusion of the paper in sec. \ref{con}.
 %%%%%%%%%%%%%%%%%%%%%%%%%%%%%%%%%%%%%%%%%%%%%%%%%%%%
\section{Formulations of Quartic Quasitopological action with nonlinear source}\label{Field}
Logarithmic (LN) and exponential nonlinear (EN) $U(1)$ gauge theories were introduced respectively, by Soleng \cite{Soleng} and Hendi \cite{Hendi1}. LN form has the ability to remove the divergence of the electric field, while EN form can reduce it. Although these forms of nonlinear
electrodynamics are not related to superstring
theory directly, but they can be shown as toy models that have the ability to produce particle-like solutions
and realize the limiting curvature hypothesis for gauge fields \cite{Soleng}. They are defined as
\begin{eqnarray}\label{non1}
\mathcal{L}(F)=\left\{
\begin{array}{ll}
$$4\beta^2[\mathrm{exp}(-\frac{F}{4\beta^2})-1]$$,\quad\quad\quad \quad\quad  \ {EN}\quad &  \\ \\
$$-8\beta^2 \mathrm{ln}[1+\frac{F}{8\beta^2}]$$,\quad\quad\quad\quad\quad\quad\quad  \ {LN}\quad &
\end{array}
\right.
\end{eqnarray}
where $F=F_{\mu\nu}F^{\mu\nu}$. $F_{\mu\nu}$ is the electromagnetic field tensor and it is defined as $F_{\mu\nu}=\partial_{\mu}A^{\nu}-\partial_{\nu}A^{\mu}$, where $A_{\mu}$ represents the vector potential. In the weak field approximation (that is described by $\beta\rightarrow\infty$), the nonlinear theory reduces to the usual linear Maxwell theory $\mathcal{L}(F)=-F_{\mu\nu}F^{\mu\nu}$. Considering the nonlinear electrodynamics theory \eqref{non1}, we start with a $(n+1)$-dimensional action in the presence of quartic quasitopological gravity
\begin{equation}\label{Act1}
I_{\rm{bulk}}=\frac{1}{16\pi}\int{d^{n+1}x\sqrt{-g}\big\{-2\Lambda+{\mathcal L}_1+\hat{\lambda} {\mathcal L}_2+\hat{\mu} {\mathcal L}_3+\hat{c}{\mathcal L}_4+\mathcal{L}(F)\big\}},
\end{equation}
which $\Lambda$ is the cosmological constant and has a negative, positive or zero value in anti-de Sitter(AdS), de Sitter(dS) or flat spacetime, respectively. ${\mathcal L}_1=R$ and ${\mathcal L}_2=R_{abcd}R^{abcd}-4R_{ab}R^{ab}+R^2$ are respectively, Einstein-Hilbert and second order Lovelock (Gauss-Bonnet) Lagrangians. ${{\mathcal L}_3}$ and ${{\mathcal L}_4}$ are also cubic and quartic quasitopological terms with definitions
\begin{eqnarray}\label{quasi3}
{{\mathcal L}_3}&=&
R_a{{}^c{{}_b{{}^d}}}R_c{{}^e{{}_d{{}^f}}}R_e{{}^a{{}_f{{}^b}}}+\frac{1}{(2n-1)(n-3)} \bigg(\frac{3(3n-5)}{8}R_{abcd}R^{abcd}R-3(n-1)R_{abcd}R^{abc}{{}_e}R^{de}\nonumber\\
&&+3(n+1)R_{abcd}R^{ac}R^{bd}+6(n-1)R_a{{}^b}R_b{{}^c}R_{c}{{}^a}-\frac{3(3n-1)}{2}R_a{{}^b}R_b{{}^a}R +\frac{3(n+1)}{8}R^3\bigg),
\end{eqnarray} and
\begin{eqnarray}\label{quasi4}
{\mathcal{L}_4}&=& c_{1}R_{abcd}R^{cdef}R^{hg}{{}_{ef}}R_{hg}{{}^{ab}}+c_{2}R_{abcd}R^{abcd}R_{ef}{{}^{ef}}+c_{3}RR_{ab}R^{ac}R_c{{}^b}+c_{4}(R_{abcd}R^{abcd})^2\nonumber\\
&&+c_{5}R_{ab}R^{ac}R_{cd}R^{db}+c_{6}RR_{abcd}R^{ac}R^{db}+c_{7}R_{abcd}R^{ac}R^{be}R^d{{}_e}+c_{8}R_{abcd}R^{acef}R^b{{}_e}R^d{{}_f}\nonumber\\
&&+c_{9}R_{abcd}R^{ac}R_{ef}R^{bedf}+c_{10}R^4+c_{11}R^2 R_{abcd}R^{abcd}+c_{12}R^2 R_{ab}R^{ab}\nonumber\\
&&+c_{13}R_{abcd}R^{abef}R_{ef}{{}^c{{}_g}}R^{dg}+c_{14}R_{abcd}R^{aecf}R_{gehf}R^{gbhd},
\end{eqnarray}
where the coefficients $c_{i}$'s are written in the appendix \eqref{app}. $\hat{\lambda}$, $\hat{\mu}$ and $\hat{c}$ are the coefficients of Gauss-Bonnet, cubic and quartic quasitopological gravities which are written as
\begin{eqnarray}
\hat{\lambda}=\frac{\lambda l^2}{(n-2)(n-3)},
\end{eqnarray}
\begin{eqnarray}
\hat{\mu}=\frac{8\mu(2n-1)l^4}{(n-2)(n-5)(3n^2-9n+4)},
\end{eqnarray}
\begin{eqnarray}
\hat{c}=\frac{cl^6}{n(n-1)(n-3)(n-7)(n-2)^2(n^5-15n^4+72n^3-156n^2+150n-42)},
\end{eqnarray}
and $l$ is a scale factor related to the cosmological constant $\Lambda$.
\section{Quartic QuasiTopological Black Hole Solutions}\label{static}
In this section, we would like to obtain the static solutions of quartic quasitopological black brane coupled to nonlinear electrodynamics. So, we begin with a $(n+1)$-dimensional static metric having a flat boundary
\begin{eqnarray}\label{metric1}
ds^2=-g(r) dt^2+\frac{d r^2}{f(r)}+r^2 \sum_{i=1}^{n-1} d\phi_{i}^2.
\end{eqnarray}
To obtain the static solutions, we consider the gauge field $A_{\mu}$ as
\begin{eqnarray}\label{vector1}
A_{\mu}=h(r)\delta_{\mu}^{0}.
\end{eqnarray}
If we substitute the relations \eqref{metric1} and \eqref{vector1} in the action \eqref{Act1} and use the redefinitions $\Psi(r)=-\frac{l^2}{r^2}f(r)$ and $g(r)=N^2(r)f(r)$, we get to an action as $I(\Psi(r),h(r), N(r))$. By varying this action with respect to $\Psi(r)$, we get to the equation
\begin{equation}\label{equ1}
\bigg(1+2\lambda \Psi(r)+3\mu \Psi^2(r)+4c\Psi^3(r)\bigg)\frac{dN(r)}{dr}=0,
\end{equation}
which shows that $N(r)$ should have a constant value and for simplicity, we choose $N(r)=1$. Then, we vary $I(\Psi(r),h(r), N(r))$ with respect to the function $h(r)$ and use $N(r)=1$, which leads to the equations
\begin{equation}\label{equ3}
\left\{
\begin{array}{ll}
$$\bigg(h^{'}r^{n-1}\mathrm {exp}\bigg[\frac{h^{'2}}{2\beta^2}\bigg]\bigg)^{'}=0$$,\quad \quad\quad\quad \quad\quad\quad\quad  \ {EN}\quad &  \\ \\
$$\bigg(h^{'}r^{n-1}(1-\frac{h^{'2}}{4\beta^2})^{-1}\bigg)^{'}=0$$.\quad\quad\quad\quad\quad\quad\quad\ {LN}\quad &
\end{array}
\right.
\end{equation}
By solving these equations, we get to the electromagnetic fields of exponential and logarithmic forms
\begin{eqnarray}\label{Fphir}
F_{tr}=-h^{'}=\left\{
\begin{array}{ll}
$$\beta\sqrt{L_{W}(\eta)}$$,\quad \quad\quad\quad \quad\quad\quad\quad\quad\quad  \ {EN}\quad &  \\ \\
$$\frac{2q}{r^{n-1}}(1+\sqrt{1+\eta})^{-1}$$,\quad\quad\quad\quad\quad\quad\quad  \ {LN}\quad &
\end{array}
\right.
\end{eqnarray}
where $\eta=\frac{q^2}{\beta^2 r^{2n-2}}$ and $q$ is the constant of integration which is related to the electric charge of the black hole. It is notable that for large $\beta$, $F_{tr}$ leads to
\begin{eqnarray}
F_{tr}=\frac{q}{r^{n-1}}-\left\{
\begin{array}{ll}
$$\frac{q^3}{2\beta^2r^{3n-3}}+\mathcal{O}(\frac{1}{\beta^4})$$,\quad \quad\quad\quad\quad\quad\quad  \ {EN}\quad &  \\ \\
$$\frac{q^3 }{4\beta^2r^{3n-3}}+\mathcal{O}(\frac{1}{\beta^4})$$,\quad\quad\quad\quad\quad\quad\quad  \ {LN}\quad &
\end{array}
\right.
\end{eqnarray}
which are the electromagnetic fields of linear Maxwell theory \cite{Naei1} plus some leading order nonlinear
correction terms.
To find the gauge potential $A_{t}$, we should solve the equation $F_{tr}+\partial_{r}A_{t}=0$ that leads to
\begin{equation}\label{hh}
\begin{split}
A_{t}=h(r)=\left\{
\begin{array}{ll}
$$\frac{n-1}{n-2}\beta\big(\frac{q}{\beta}\big)^{\frac{1}{n-1}}\big(L_{W}(\eta)\big)^{\frac{n-2}{2(n-1)}} {}_2F_{1}\bigg(\big[\frac{n-2}{2(n-1)}\big]\,,\big[\frac{3n-4}{2(n-1)}\big]\,,-\frac{1}{2(n-1)}L_{W}(\eta)\bigg)
-\beta r \sqrt{L_{W}(\eta)}$$,\quad\quad\quad  \ {EN}\quad &  \\ \\
$$\frac{q}{(n-2)r^{n-2}}{}_{3}F_{2}([\frac{n-2}{2(n-1)},\frac{1}{2},1]\,,[\frac{3n-4}{2(n-1)},2]\,,-\eta)$$,\quad\quad\quad\quad\quad\quad\quad\quad\quad\quad\quad\quad\quad\quad\quad\quad\quad\quad \quad \quad\quad  \ {LN}\quad &
\end{array}
\right.
\end{split}
\end{equation}
where $L_{W}$ is the Lambert function which obeys the relation $L_{W}(x)e^{L_{W}(x)}=x$ and ${}_2F_{1}([a],[b],c)$ and ${}_3F_{2}([d,e,f],[g,h],i)$ are the hypergeometric functions. As $r\rightarrow\infty$, $A_{t}$ reduces to
\begin{eqnarray}
A_{t}=\frac{q}{(n-2)r^{n-2}}+\mathcal{O}\bigg(\frac{1}{\beta^2}\bigg),
\end{eqnarray}
which describes the vector potential of Maxwell theory \cite{Naei1}. At last, if we vary the action $I(\Psi(r),h(r), N(r))$ with respect to the function $N(r)$ and put $N(r)=1$ and Eq. \eqref{Fphir} in it, we get to equation
\begin{eqnarray}\label{eq3}
\hat{\mu}_{4} \Psi^4+\hat{\mu}_{3} \Psi^3+\hat{\mu}
_{2} \Psi^2+\Psi+\kappa=0,
\end{eqnarray}
where
\begin{eqnarray}\label{kappa1}
\kappa&=&-\frac{2\Lambda l^2}{n(n-1)}-\frac{m}{(n-1)r^n}+\nonumber\\
&&\left\{
\begin{array}{ll}
$$-\frac{4 l^2\beta^2}{n(n-1)}+\frac{4(n-1)\beta q l^{2}}{n(n-2)r^n}(\frac{q}{\beta})^{\frac{1}{n-1}}(L_{W}(\eta))^{\frac{n-2}{2(n-1)}}\times{}_2 F_{1}([\frac{n-2}{2(n-1)}]\,,[\frac{3n-4}{2(n-1)}]\,,-\frac{1}{2(n-1)}L_{W}(\eta))\\-\frac{4\beta q l^{2}}{(n-1)r^{n-1}}[L_{W}(\eta)]^{\frac{1}{2}}\times[1-\frac{1}{n}(L_{W}(\eta))^{-1}]$$,\quad \quad\quad\quad\quad\quad \quad\quad\quad\quad\quad\quad\quad  \ {EN}\quad &  \\ \\
$$\frac{8(2n-1)}{n^2(n-1)}\beta^2 l^2[1-\sqrt{1+\eta}]+\frac{8(n-1)q^2l^{2}}{n^2 (n-2)r^{2n-2}}{}_2 F_{1}([\frac{n-2}{2(n-1)},\frac{1}{2}]\,,[\frac{3n-4}{2(n-1)}]\,,-\eta)\\
-\frac{8}{n(n-1)}l^2\beta^2 \mathrm {ln}[\frac{2\sqrt{1+\eta}-2}{\eta}]$$,\quad\quad\quad\quad\quad\quad\quad\quad\quad\quad\quad\quad\quad\quad\quad\quad\quad\quad \quad\quad  \ {LN}\quad &
\end{array}
\right.
\end{eqnarray}
and $m$ is a constant of integration that is related to the mass of the black hole. Eq. \eqref{eq3} leads to real solutions, if the condition
\begin{eqnarray}
{\hat{\mu}_{2}}^{2}<3\hat{\mu}_{3}-12\hat{\mu}_{4} \kappa,
\end{eqnarray}
is satisfied. So by this condition, the function $f(r)$ may be obtained as
\begin{eqnarray}\label{func1}
f(r)=\frac{-r^2}{l^2}\bigg(-\frac{\mu}{4c}+\frac{\pm_{s}W\mp_{t}\sqrt{-(3\alpha+2y\pm_{s}\frac{2\beta}{W})}}{2}\bigg),
\end{eqnarray}
where $W$, $\alpha$, $y$ and $\beta$ are introduced in appendix \eqref{app1} and two $\pm_{s}$ should have both the same sign, while the sign of $\pm_{t}$ is independent.

\section{Quartic QuasiTopological Rotating Black Brane Solutions}\label{rotating}
Now in this section, we would like to extend our static solutions to the rotating ones by transformation
\begin{eqnarray}\label{trans}
t^{'}=\Xi t-\sum_{i=1}^{k}a_{i}\phi_{i}\,\,\,\,,\,\,\,\,\,\,\,\phi_{i}^{'}=\frac{a_{i}}{l^2}t-\Xi\phi_{i}.
\end{eqnarray}
in the metric \eqref{metric1}. For this purpose, we consider the rotation group $SO(n)$ in $(n+1)$ dimensions. The maximum number of independent rotation parameters is equal to the number of Casimir operators which is $[n/2]$ and $[x]$ is the integer part of $x$. So, the metric of $(n+1)$-dimensional rotating spacetime with $k\leq[n/2]$ rotation parameters with flat horizon can be written as
\begin{eqnarray}\label{metric2}
ds^2&=&-N^2(r)f(r)\bigg(\Xi dt-\sum_{i=1}^{k}a_{i} d\phi_{i}\bigg)^2+\frac{r^2}{l^4}\sum_{i=1}^{k}(a_{i} dt-\Xi l^2 d\phi_{i})^2-\frac{r^2}{l^2}\sum_{i<j}^{k}(a_{i}d\phi_{j}-a_{j}d\phi_{i})^2\nonumber\\&&+\frac{d r^2}{f(r)}+r^2 \sum_{i=k+1}^{n-1} d\phi_{i}^2,\\
\Xi&=& \sqrt{1+\sum_{i=1}^{k}a_{i}^2/l^2},
\end{eqnarray}
where $a_{i}$'s are $k$ rotation parameters. It is clear that the metrics \eqref{metric1} and \eqref{metric2} can be mapped to each other by transformations \eqref{trans} locally, not globally. Also, the vector potential for rotating solutions is defined as
\begin{eqnarray}
A_{\mu}=h(r)(\Xi dt-\sum_{i=1}^{k}a_{i}d\phi_{i}),
\end{eqnarray}
where $h(r)$ and $f(r)$ are respectively the same as the ones in equations \eqref{hh} and \eqref{func1}. It is also necessary to mention that choosing the value $a_{i}=0$ (or $\Xi=1$) in the above relations, will reach us to the static solutions in the previous section. So, from here onwards, for economic reasons, we investigate the behavior of the rotating solutions that can be generalized to the static ones by choosing $a_{i}=0$ (or $\Xi=1$).\\

\section{physical structure of the rotating black brane}\label{structure}
\begin{figure}
\center
\includegraphics[scale=0.5]{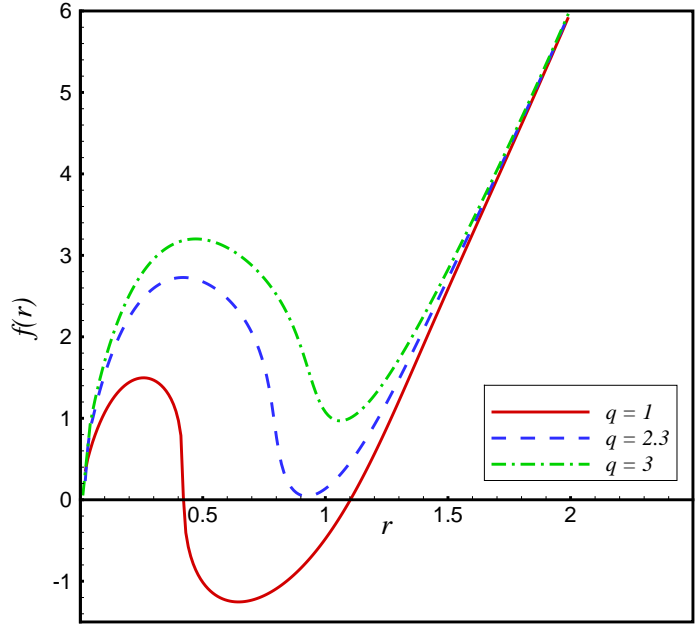}
\caption{\small{AdS solution $f(r)$ versus $r$ for different $q$, with $m=7$, $n=5$, $\beta=7$, ${\hat{\mu}}_{2}=0.4$, ${\hat{\mu}}_{3}=0.1$ and ${\hat{\mu}}_{4}=-0.002$.} \label{figure1}}
\end{figure}

\begin{figure}
\center
\includegraphics[scale=0.5]{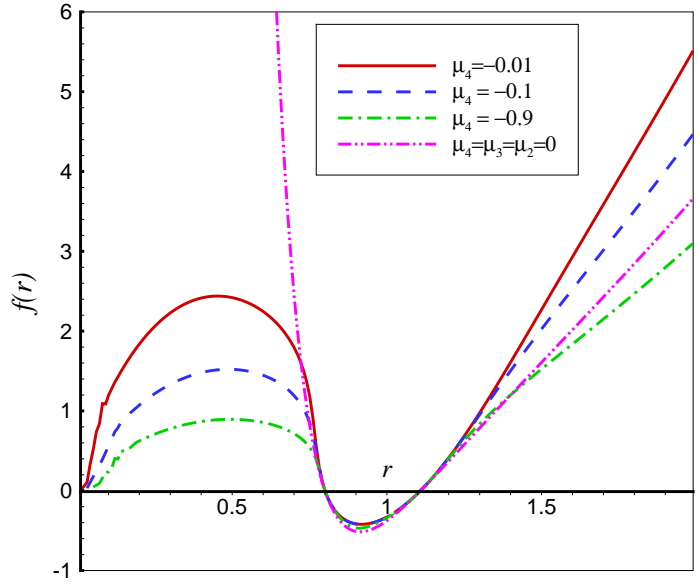}
\caption{\small{AdS solution $f(r)$ versus $r$ with $m=10$, $n=5$, $\beta=10$, $q=2.6$ and ${\hat{\mu}}_{2}=0.4$ and ${\hat{\mu}}_{3}=0.1$ for three solid red, dash blue and dash-dot green diagrams and ${\hat{\mu}}_{2}=0$ and ${\hat{\mu}}_{3}=0$ for a dash-dot-dot pink diagram.} \label{figure2}}
\end{figure}

\begin{figure}
\center
\includegraphics[scale=0.5]{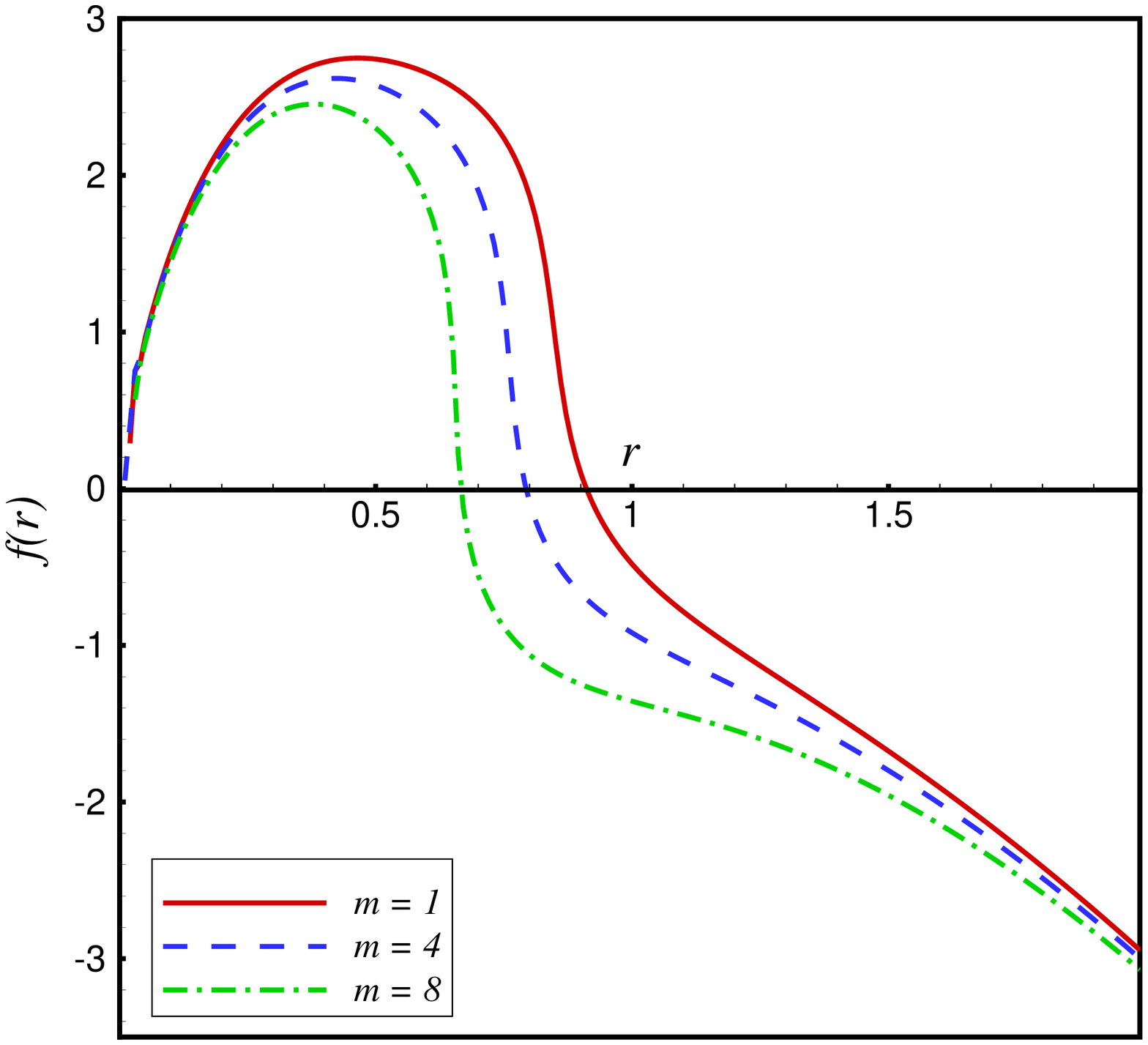}
\caption{\small{dS solution $f(r)$ versus $r$ for different $m$, with $n=5$, $\beta=7$, $q=2$, ${\hat{\mu}}_{2}=0.4$, ${\hat{\mu}}_{3}=0.1$ and ${\hat{\mu}}_{4}=-0.002$.} \label{figure3}}
\end{figure}

Now, we tend to have a study on physical structures of the obtained solutions in quartic quasitopological gravity with nonlinear electrodynamics. For this purpose, we probe behavior of Kretschmann scalar which goes to infinity as $r$ tends to zero. This suggests that both static and rotating spacetimes have an essential singularity located at $r=0$.\\
To know more about the function $f(r)$, we have plotted $f(r)$ versus $r$ in Figs. \ref{figure1}-\ref{figure3}. We should say that as plots of functions in LN form are similar to the ones in EN form, so we use LN form to plot the figures of this paper. For simplicity, we have considered $l=1$. In all three figures, the function $f(r)$ has a zero value at $r=0$, while for $r\rightarrow\infty$, it depends on the sign of $\Lambda$. In this limit, $f(r)$ goes to $+\infty$, $0$ or $-\infty$ if it is in AdS($\Lambda<0$), flat($\Lambda=0$)  or AdS($\Lambda>0$) spacetime, respectively. In Fig. \ref{figure1}, we have plotted $f(r)$ versus $r$ for different values of $q$ in AdS spacetime. This figure shows that for fixed values of the parameters $m$, $n$, $\beta$ and $\hat{\mu}_{i=2,3,4}$, there is a $q_{\rm{ext}}$ for which we have an extreme black hole/brane while for $q<q_{\rm{ext}}$, we have a black hole/brane with two horizons and for $q>q_{\rm{ext}}$, there is a naked singularity. \\
In Fig. \ref{figure2}, we have compared the behavior of $f(r)$ in quasitopological gravity(the three solid red, dash blue and dash-dot green diagrams) with the one in Einstein's gravity(a dash-dot-dot pink diagram). It can be seen that all four diagrams show different black holes but with the same horizons. This important point is also clear in relation \eqref{eq3}, which shows that by choosing $f(r_{+})=0$ in this equation, the value of the horizons are independent to the value of $\hat{\mu}_{4}$ and therefore of the kind of gravity. We can also see that unlike the behavior of $f(r)$ in quasitopological gravity, for fixed parameters, this function goes to infinity at the origin in Einstein's gravity. This can be the priority of quasitopological gravity to Einstein's gravity.\\
In Fig. \ref{figure3}, we have plotted $f(r)$ versus $r$ for different values of mass parameter $m$ in dS spacetime. We can see that for fixed values of parameters $n$, $\beta$, $q$ and $\hat{\mu}_{i=2,3,4}$, the value of the horizon $r_{+}$ decreases, as the value of $m$ increases. So, the larger the mass of a black hole, the smaller the size of its horizon.\\
To know the other prpperties of these solutions like angular velocity, electric potential and temperature, we should note the killing vector of the rotating black brane solutions
\begin{eqnarray}
\chi=\partial_{t}+\sum_{i=1}^{k}\Omega_{i}\partial_{\phi_{i}},
\end{eqnarray}
where $\Omega_{i}$ is the angular velocity of the horizon defined as
\begin{eqnarray}\label{Omega}
\Omega_{i}=-\bigg(\frac{g_{t\phi_{i}}}{g_{\phi_{i}\phi_{i}}}\bigg)_{r=r_{+}}=\frac{a_{i}}{\Xi l^2}.
\end{eqnarray}
The electric potential $\Phi$ at infinity with respect to the horizon is also defined as
\begin{eqnarray}
\Phi=A_{\mu}\chi^{\mu}|_{r\rightarrow\infty}-A_{\mu}\chi^{\mu}|_{r=r_{+}},
\end{eqnarray}
\begin{equation}\label{A1}
\begin{split}
\Phi=\left\{
\begin{array}{ll}
$$\frac{n-1}{n-2}\frac{\beta}{\Xi}\big(\frac{q}{\beta}\big)^{\frac{1}{n-1}}\big(L_{W}(\eta_{+})\big)^{\frac{n-2}{2(n-1)}} {}_2F_{1}\bigg(\big[\frac{n-2}{2(n-1)}\big]\,,\big[\frac{3n-4}{2(n-1)}\big]\,,-\frac{1}{2(n-1)}L_{W_{+}}(\eta)\bigg)
-\beta r_{+} \sqrt{L_{W}(\eta_{+})}$$,\quad\quad\quad  \ {EN}\quad &  \\ \\
$$\frac{q}{\Xi(n-2)r_{+}^{n-2}}{}_{3}F_{2}([\frac{n-2}{2(n-1)},\frac{1}{2},1]\,,[\frac{3n-4}{2(n-1)},2]\,,-\eta_{+})$$.\quad\quad\quad\quad\quad\quad\quad\quad\quad\quad\quad\quad\quad\quad\quad\quad\quad\quad \quad \quad\quad  \ {LN}\quad &
\end{array}
\right.
\end{split}
\end{equation}
We can obtain the Hawking temperature of this rotating black brane on the outer
horizon $r_{+}$ through the use of surface gravity $\kappa$ as
\begin{eqnarray}\label{Temp}
T=\frac{\kappa}{2\pi\Xi}=\frac{1}{2\pi\Xi}\sqrt{-\frac{1}{2}(\nabla_{\mu}\chi_{\nu})(\nabla^{\mu}\chi^{\nu})}=\frac{f^{'}(r_{+})}{4\pi\Xi}=\frac{r^2}{4\pi \Xi l^2}\kappa^{'},
\end{eqnarray}
where the primes represent the derivative with respect to $r$.

\section{Thermodynamics of the rotating black brane}\label{Thermo}
In this section, we are going to obtain thermodynamic properties of the solutions. Varying the action \eqref{Act1} with respect to the metric to gain thermodynamic quantities, one faces with a total derivative that has a surface integral involving
the derivative of $\delta g_{\mu\nu}$ normal to the boundary. This makes the variation of the action ill defined because the normal derivative terms can not cancel each other. To solve this problem, we should add Gibbons-Hawking surface term $I_{b}$ to the bulk action \eqref{Act1}.
$I_{b}$ makes the variational
principle well defined if we choose it as
\begin{eqnarray}
I_{b}=I_{b}^{(1)}+I_{b}^{(2)}+I_{b}^{(3)}+I_{b}^{(4)},
\end{eqnarray}
where $I_{b}^{(1)}$, $I_{b}^{(2)}$, $I_{b}^{(3)}$ and $I_{b}^{(4)}$ are respectively, the proper surface terms for Hilbert-Einstein \cite{Gibbons}, Gauss-Bonnet\cite{Myers,Davis}, third order \cite{Vahid1} and fourth order quasitopological \cite{Bazr2} gravities that are obtained as
\begin{eqnarray}
I_{b}^{(1)}=\frac{1}{8\pi}\int_{\partial\mathcal{M}} d^{n}x \sqrt{-\gamma}K,
\end{eqnarray}
\begin{eqnarray}
I_{b}^{(2)}=\frac{1}{8\pi}\int_{\partial\mathcal{M}} d^{n}x \sqrt{-\gamma}\frac{2{\hat{\mu}}_{2}l^2}{3(n-2)(n-3)}(3KK_{ac}K^{ac}-2K_{ac}K^{cd}K_{d}^{a}-K^3),
\end{eqnarray}
\begin{eqnarray}
I_{b}^{(3)}&=&\frac{1}{8\pi}\int_{\partial\mathcal{M}} d^{n} x \sqrt{-\gamma}\bigg\{\frac{3\hat{\mu}_{3}l^4}{5n(n-2)(n-1)^2 (n-5)}(nK^5-2K^3K_{ab}K^{ab}+4(n-1)K_{ab}K^{ab}K_{cd}K^{d}_{e}K^{ec}\nonumber\\&&-(5n-6)K K_{ab}[nK^{ab}K^{cd}K_{cd}-(n-1)K^{ac}K^{bd}K_{cd}])\bigg\},
\end{eqnarray}
\begin{eqnarray}
I_{b}^{(4)}&=&\frac{1}{8\pi}\int_{\partial\mathcal{M}} d^{n} x \sqrt{-\gamma}\frac{2\hat{\mu}_{4}l^6}{7n(n-1)(n-2)(n-7)(n^2-3n+3)}\bigg\{\alpha_{1}K^3K^{ab}K_{ac}K_{bd}K^{cd}+\alpha_{2}K^2K^{ab}K_{ab}K^{cd}K^{e}_{c}K_{de}\nonumber\\
&&+\alpha_{3}K^2K^{ab}K_{ac}K_{bd}K6{ce}K^{d}_{e}+\alpha_{4} K K^{ab}K_{ab}K^{cd}K^{e}_{c} K^{f}_{d}K_{ef}+\alpha_{5} K K^{ab} K^{c}_{a}K_{bc}K^{de}K^{f}_{d}K_{ef}+\alpha_{6} K K^{ab}K_{ac} K_{bd} \nonumber\\
&& K^{ce}K^{df}K_{ef}+\alpha_{7} K^{ab}K^{c}_{a}K_{bc}K^{de}K_{df}K_{eg}K^{fg}\bigg\},
\end{eqnarray}
where $\gamma_{ab}$ is the induced metric on the boundary $\partial\mathcal{M}$ and $K$ is the trace of extrinsic curvature $K^{ab}$ of this boundary.
The value of the total action $I_{\rm{bulk}}+I_{b}$ is divergent on solutions. Using the counterterm method inspired by AdS/CFT correspondence, we can add a counterterm action $I_{ct}$ to remove this divergence \cite{Brown,Nojiri}. This should have a form like
\begin{eqnarray}
I_{ct}= -\frac{1}{8\pi}\int_{\partial\mathcal{M}} d^{n} x\sqrt{-\gamma}\frac{n-1}{L_{\rm{eff}}},
\end{eqnarray}
where $L_{\rm{eff}}$ is a scale length factor and reduces
to $l$ as $\hat{\mu}_{2}$, $\hat{\mu}_{3}$ and $\hat{\mu}_{4}\rightarrow 0$. It should be defined as
\begin{eqnarray}
L_{\rm{eff}}=-\frac{210\Psi_{\infty}^{1/2}}{15\hat{\mu}_{4}\Psi_{\infty}^{4}+21\hat{\mu}_{3}\Psi_{\infty}^{3}+35\hat{\mu}_{2}\Psi_{\infty}^{2}-105\Psi_{\infty}-105}l,
\end{eqnarray}
where $\Psi_{\infty}$ is the limit of $\Psi$ at infinity in Eq. \eqref{eq3}. To calculate the conserved quantities, we should choose a spacelike surface $\mathcal{B}$ in $\partial{\mathcal{M}}$ with metric $\sigma_{ij}$, and write the boundary metric in ADM (Arnowitt-Deser-Misner) form:
\begin{eqnarray}
\gamma_{ab}dx^{a}dx^{b}=-N^2dt^2+\sigma_{ij}(d\phi^{i}+V^{i}dt)(d\phi^{j}+V^{j}dt),
\end{eqnarray}
where the coordinates $\phi^{i}$ are the angular variables parameterizing the hypersurface of constant $r$
around the origin, and $N$ and $V_{i}$ are the lapse and shift functions, respectively. If we evaluate the finite stress tensor $T_{ab}$ by the new finite action $I_{\rm{bulk}}+I_{b}+I_{ct}$ and consider a killing vector field $\xi$ on the boundary, the conserved quantities are obtained as
\begin{eqnarray}
\mathcal{Q}=\int_{\mathcal{B}}d^{n-1} \phi \sqrt{\sigma} T_{ab} n ^{a} \xi^{b},
\end{eqnarray}
where $\sigma$ is the determinant of the metric $\sigma_{ij}$ and $n^{a}$ is the unit normal vector on the boundary $\mathcal{B}$. Considering the boundaries with timelike $(\xi=\partial/\partial t)$ and rotational
($\varsigma=\partial/\partial  \phi$) Killing vector fields, as the boundary $\mathcal{B}$ goes to infinity, the mass and the angular momentum per unit volume $V_{n-1}$ of this black brane are obtained as
\begin{eqnarray}\label{mass}
M=\int_{\mathcal{B}} d^{n-1}\phi \sqrt{\sigma}T_{ab}n^{a}\xi^{b}=\frac{(n\Xi^2-1)}{16\pi(n-1)l^{n-1}}m,
\end{eqnarray}
\begin{eqnarray}\label{angular}
J_{i}=\int_{\mathcal{B}} d^{n-1}\phi \sqrt{\sigma}T_{ab}n^{a}\varsigma_{i}^{b}=\frac{1}{16\pi(n-1)l^{n-1}}n\Xi m a_{i}.
\end{eqnarray}
It is clear that for $a_{i}=0$ (or $\Xi=1$), the angular momentum $J_{i}$ vanishes and we get to the mass of the static black hole. To calculate the electric charge of this brane, we first consider the projections of the electromagnetic field tensor on special hypersurfaces. The normal to these hypersurfaces is
\begin{eqnarray}
u^{0}=\frac{1}{N},\,\,\,\,u^{r}=0,\,\,\,\,u^{i}=-\frac{V^{i}}{N},
\end{eqnarray}
and the electric field is $E^{\mu}=g^{\mu\rho}F_{\rho \nu}u^{\nu}$. Calculating the flux of the electric field at infinity, the electric charge per unit volume $V_{n-1}$ is obtained as
\begin{eqnarray}\label{charge}
Q=\frac{\Xi q}{4\pi l^{n-3}}.
\end{eqnarray}
In the so-called area law of entropy, the entropy is a quarter of the event horizon area \cite{beke}. Using this, the entropy per unit volume $V_{n-1}$ for this black brane is obtained as
\begin{eqnarray}\label{entropy}
S=\frac{1}{4l^{n-3}}\Xi r_{+}^{(n-1)}.
\end{eqnarray}
Now, we want to verify the first law of thermodynamics. For this purpose, we should obtain the mass $M$ as a function of extensive quantities $S$, $Q$ and $J$. Using $Z=\Xi^2$ and considering the relations \eqref{mass} and \eqref{angular}, we can get to a Smarr-type formula
\begin{eqnarray}\label{Msmar}
M(S,Q,J)=\frac{[nZ-1]}{nl\sqrt{Z(Z-1)}}J,
\end{eqnarray}
where manifests that the parameter $Z$ should be a function of the extensive parameters. We can use relations \eqref{charge} and \eqref{entropy} and the fact that $f(r_{+})=0$, in Eq. \eqref{angular} and get to an equation
$E(S,Q,J)=0$, which helps us to obtain $\big(\frac{\partial Z}{\partial X_{i}}\big)$, and $X_{i}=S,Q,J$. For example,
\begin{eqnarray}
\bigg(\frac{\partial Z}{\partial S}\bigg)_{Q,J}=-\frac{\bigg(\frac{\partial E(S,Q,J)}{\partial S}\bigg)_{Q,J}}{\bigg(\frac{\partial E(S,Q,J)}{\partial Z}\bigg)_{Q,J}},
\end{eqnarray}
which is usable for evaluating
\begin{eqnarray}\label{relation}
\bigg(\frac{\partial M}{\partial S}\bigg)_{Q,J}=\bigg(\frac{\partial M}{\partial Z}\bigg)_{Q,J}\bigg(\frac{\partial Z}{\partial S}\bigg)_{Q,J}.
\end{eqnarray}
Therefore, with these relations, we can obtain the intensive parameters $T$, $\Phi$ and $\Omega_{i}$ related to $S$, $Q$, and $J_{i}$ by
\begin{eqnarray}
T=\bigg(\frac{\partial M}{\partial S}\bigg)_{Q,J},\,\,\,\Phi=\bigg(\frac{\partial M}{\partial Q}\bigg)_{S,J},\,\,\,\,\,\,\Omega_{i}=\bigg(\frac{\partial M}{\partial J_{i}}\bigg)_{S,Q}.
\end{eqnarray}
Our calculations show that the obtained intensive parameters are the same as the results in equations \eqref{Temp}, \eqref{A1} and \eqref{Omega}. So, our obtained solutions obey the first law of thermodynamics as
\begin{eqnarray}
dM=TdS+\sum_{i=1}^{k}\Omega_{i}dJ_{i}+UdQ.
\end{eqnarray}

\section{Thermal stability in grand canonical ensemble}\label{stability}
In this section, we would like to study thermal stability of the solutions. Generally, we can probe thermal stability of a black hole as a thermodynamic system by investigating the behavior of energy $M(S,Q,J)$ with respect to small variations of thermodynamic coordinates $S$, $Q$ and $J$. To have a local stability, $M(S,Q,J)$ should be a convex function of its extensive variables. In grand canonical ensemble, positive values of Hessian matrix's determinant and temperature guarantee the stability of the solutions. The Hessian matrix of our solutions is defined as
\begin{eqnarray}
H=\left[
\begin{array}{ccc}
\frac{\partial^2 M}{\partial S^2} & \frac{\partial^2 M}{\partial S\partial Q} & \frac{\partial^2 M}{\partial S \partial J} \\
\frac{\partial^2 M}{\partial S \partial Q}& \frac{\partial^2 M}{\partial Q^2}& \frac{\partial^2 M}{\partial Q \partial J} \\
\frac{\partial^2 M}{\partial J \partial S} & \frac{\partial^2 M}{\partial Q \partial J}&\frac{\partial^2 M}{\partial J^2}
\end{array} \right],
\end{eqnarray}
in which, we have used the relation $\frac{\partial^2 M}{\partial X \partial Y}=\frac{\partial^2 M}{\partial Y \partial X}$. To find the arrays of the above matrix, we can get help from Eqs. \eqref{Msmar}-\eqref{relation}.\\
\begin{figure}
\centering
\subfigure[AdS solution]{\includegraphics[scale=0.27]{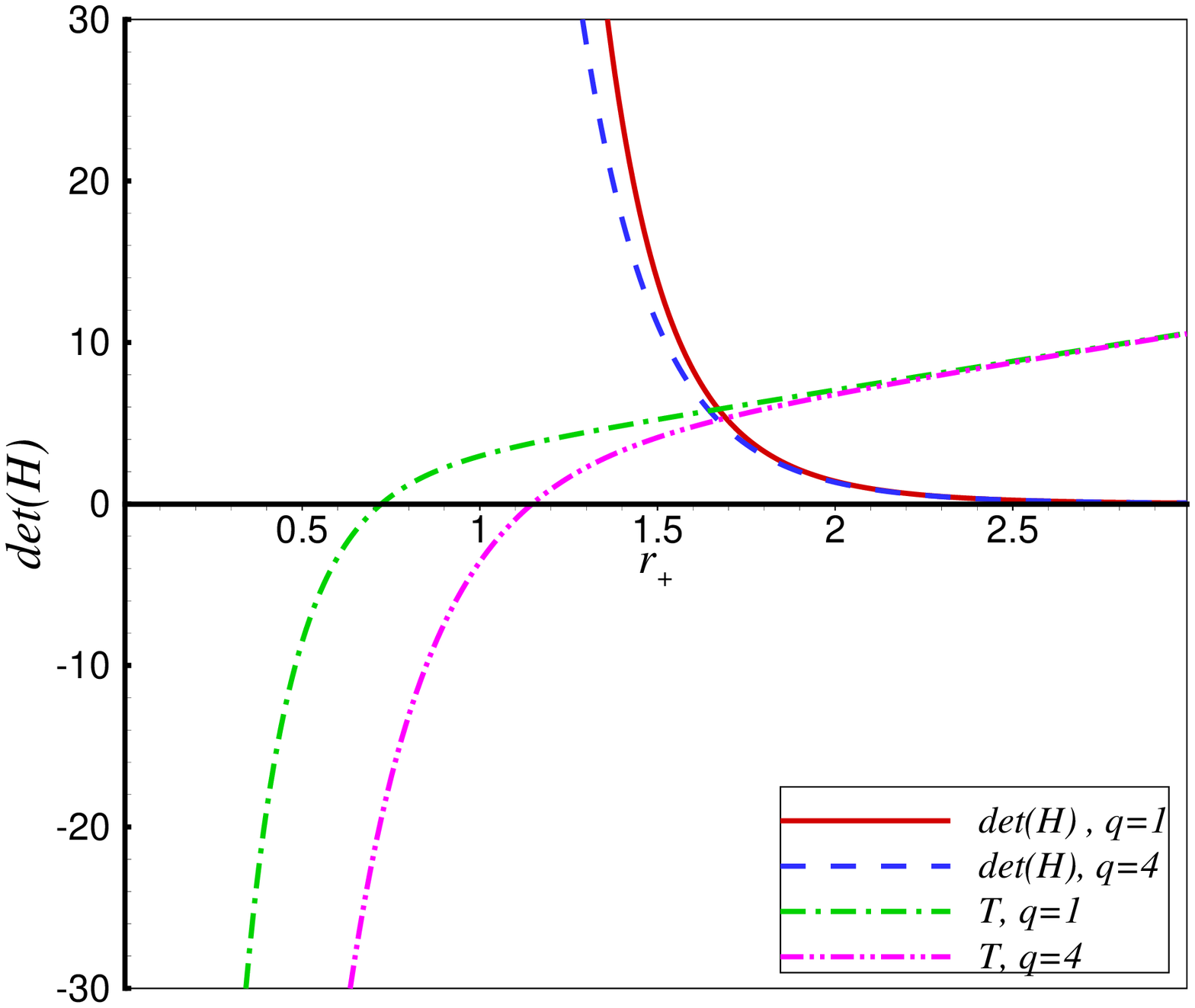}\label{fig4a}}\hspace*{.2cm}
\subfigure[dS solution]{\includegraphics[scale=0.27]{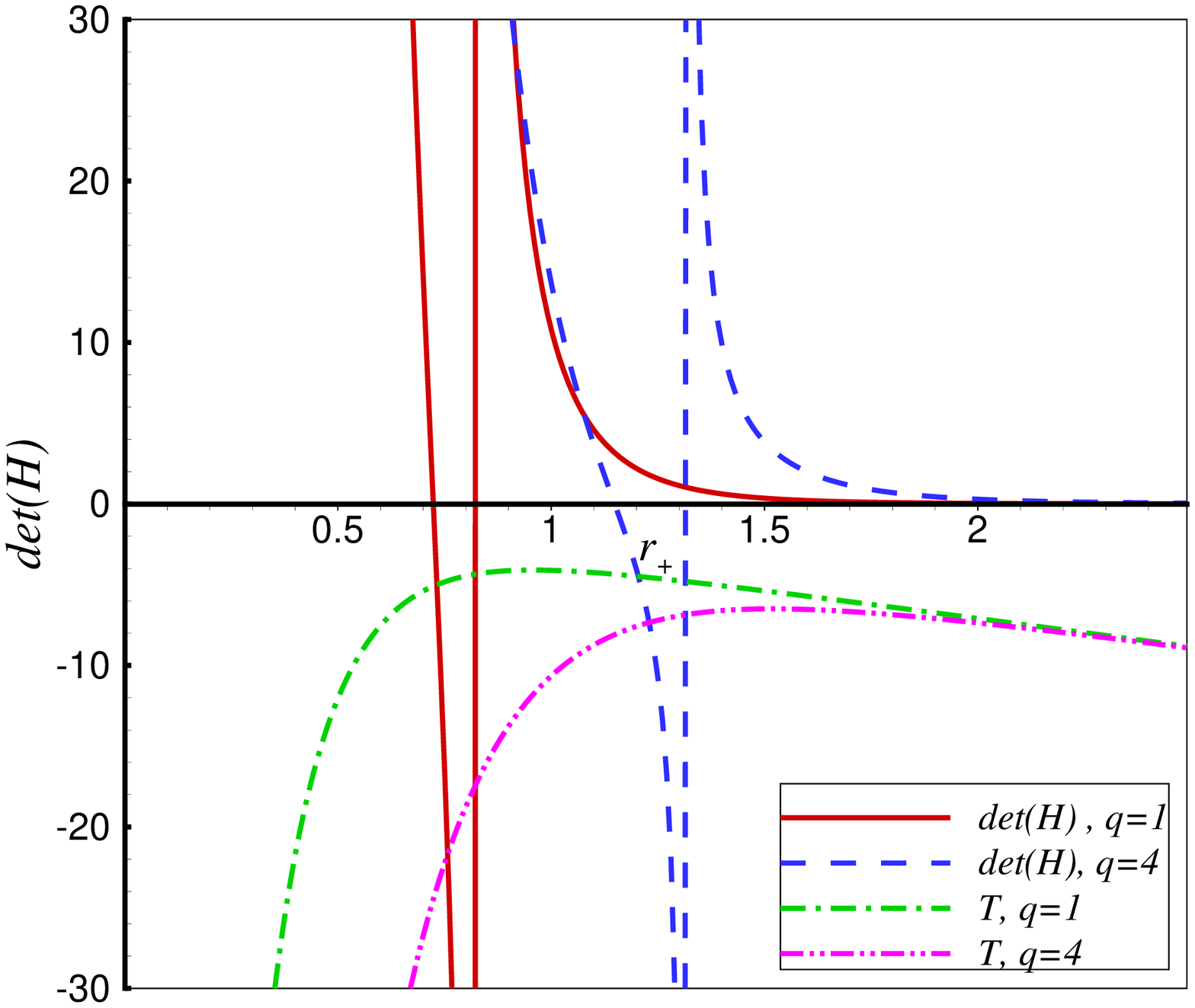}\label{fig4b}}\hspace*{.2cm}
\subfigure[flat solution]{\includegraphics[scale=0.27]{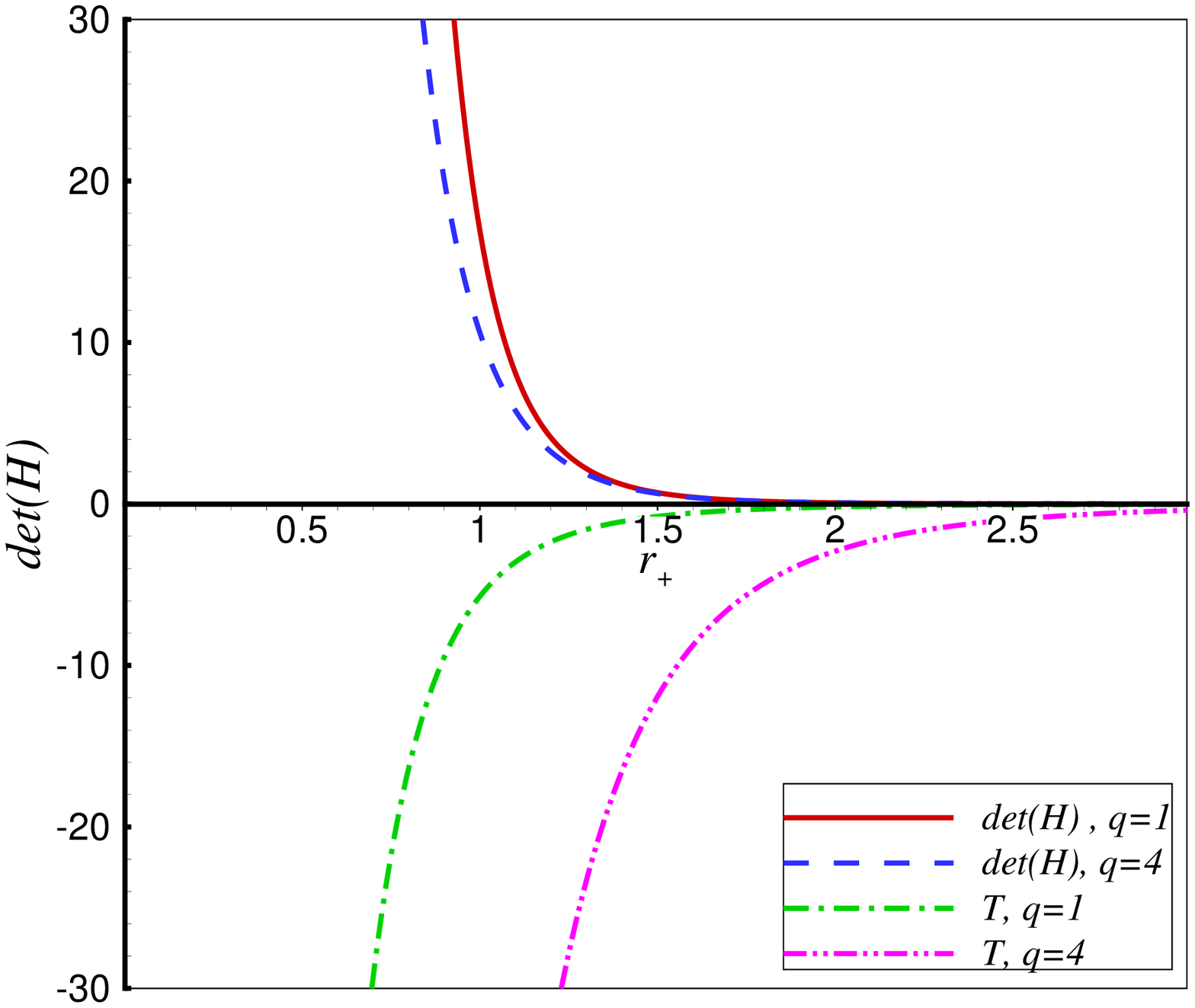}\label{fig4c}}\caption{$det(H)$ and $T$ versus $r_{+}$ for different $q$ with $\beta=2$, $\Xi=0.9$ and $n=4$.}\label{figure4}
\end{figure}
\begin{figure}
\centering
\subfigure[AdS solution]{\includegraphics[scale=0.27]{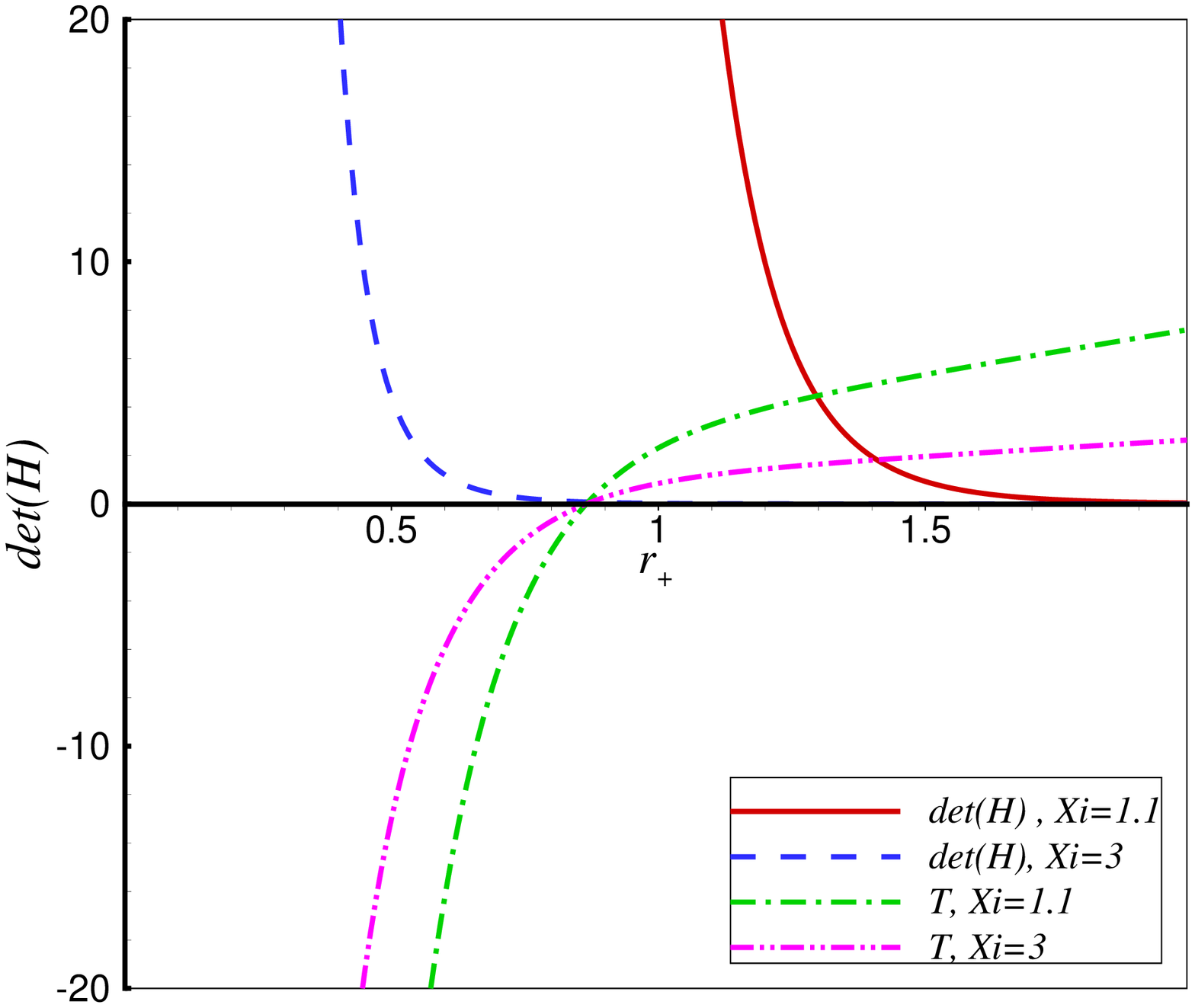}\label{fig5a}}\hspace*{.2cm}
\subfigure[dS solution]{\includegraphics[scale=0.27]{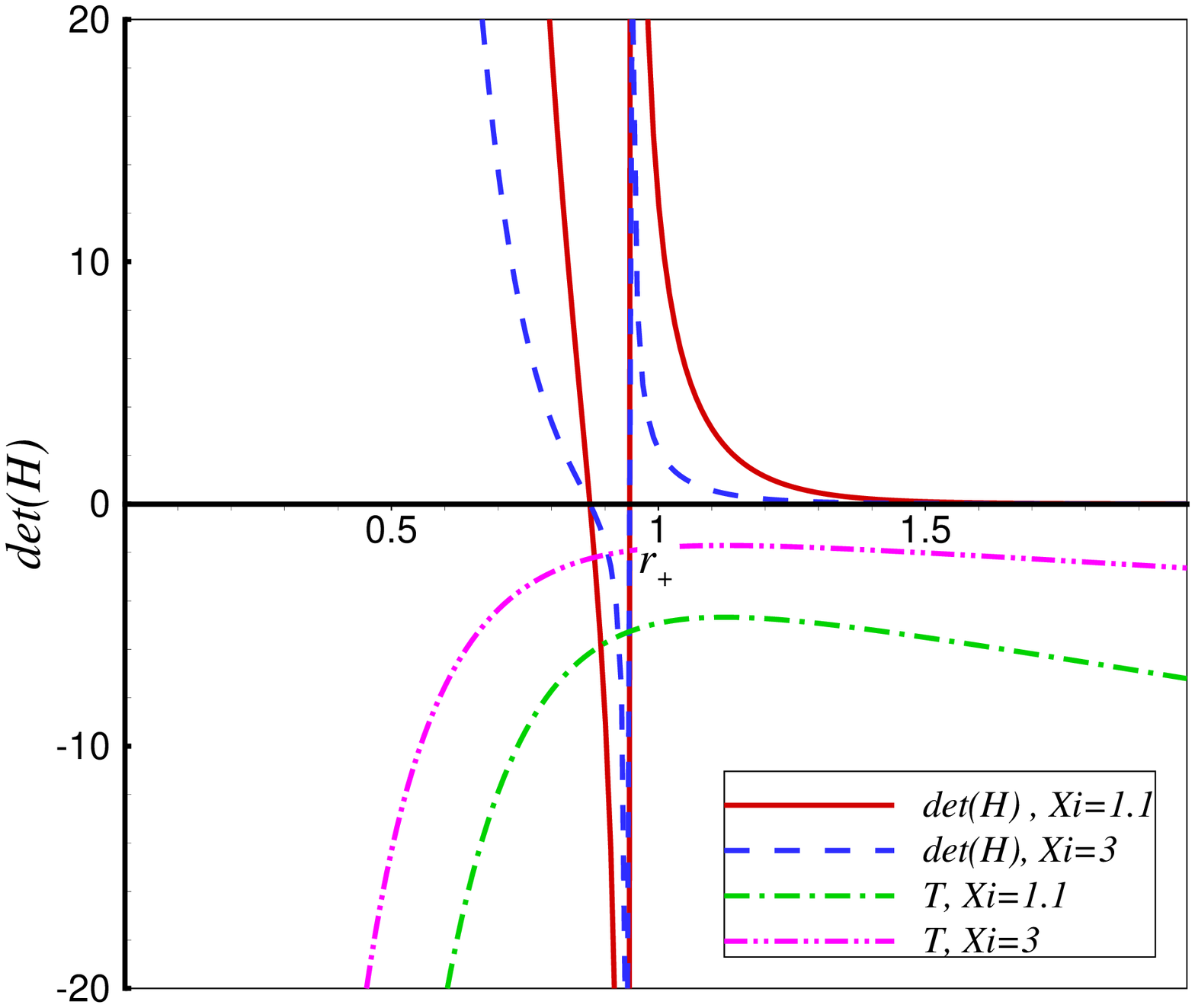}\label{fig5b}}\hspace*{.2cm}
\subfigure[flat solution]{\includegraphics[scale=0.27]{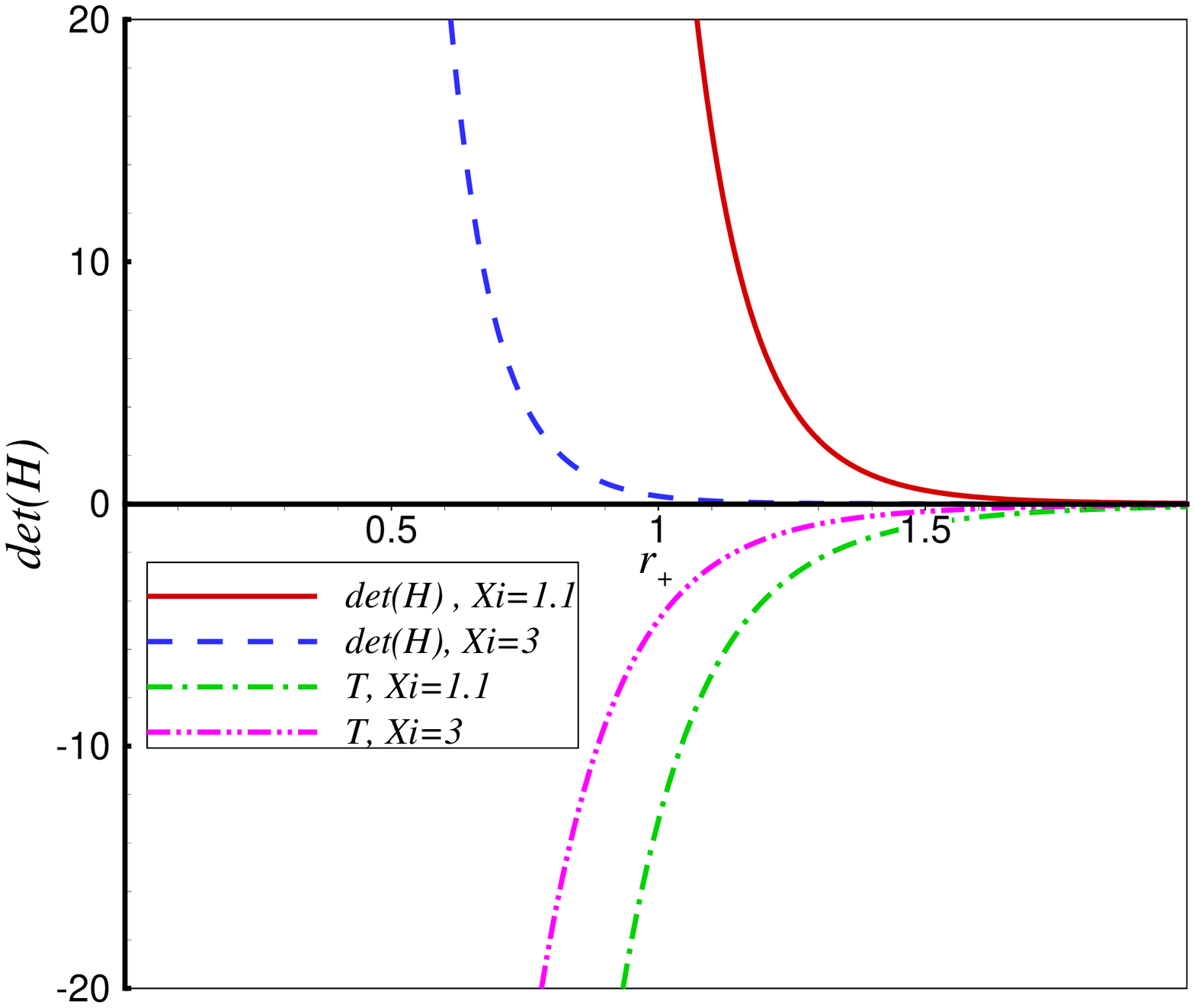}\label{fig5c}}\caption{$det(H)$ and $T$ versus $r_{+}$ for different $\Xi$ with $\beta=2$, $q=2$ and $n=5$.}\label{figure5}
\end{figure}
To peruse the stability of nonlinear quartic quasitopological black brane, we have plotted $det(H)$(we have abbreviated determinant of the Hessian's matrix) and temperature(T) versus $r_{+}$ in Figs. \ref{figure4} and \ref{figure5}. In Fig.\ref{figure4}, we have investigated the stability of our solutions for different values of charge $q$ in AdS, dS and flat spacetimes, in Figs. \ref{fig4a}, \ref{fig4b} and \ref{fig4c}, respectively. It is clearly seen that for fixed values of parameters $\beta$, $\Xi$ and $n$, T is negative for each values of $r_{+}$ in dS and flat spacetimes. So, for these parameters, dS and flat solutions are not thermally stable. For AdS solutions, $det(H)$ is positive for each values of $r_{+}$ in Fig. \ref{fig4a} and the condition of stability depends on the behavior of temperature. This figure shows that there is a ${r_{+}}_{\rm{ext}}$ that for $r_{+}>{r_{+}}_{\rm{ext}}$, $T>0$ and T is negative, for $r_{+}<{r_{+}}_{\rm{ext}}$. The value of ${r_{+}}_{\rm{ext}}$ becomes larger as the charge parameter $q$ increases.\\
In Fig. \ref{figure5}, the stability of our solutions for different values of $\Xi$ in AdS, dS and flat spacetimes is under investigation. Again, it is obvious that $T<0$, for dS and flat spacetimes. So, by this, we can conclude totally that our nonlinear quartic quasitopological black brane is thermally unstable in dS and flat spacetimes. In fig. \ref{fig5a}, as $det(H)$ is positive for each values of $r_{+}$, therefore, the stability is related to the sign of $T$.  Again, we have also a ${r_{+}}_{\rm{ext}}$ that $r_{+}>{r_{+}}_{\rm{ext}}$ is a acceptable region for stability. This figure also manifests that the value of ${r_{+}}_{\rm{ext}}$ doesn't depend on the value of $\Xi$. This issue is also clear in Eq. \ref{Temp}. In order to have a zero value for temperature, the value of $\kappa$ should be $0$ where it is not dependent on the values of the parameter $\Xi$. \\

\section{Geometrothermodynamics}\label{geom}
\begin{figure}
\centering
\subfigure[$\Xi=0.9$]{\includegraphics[scale=0.27]{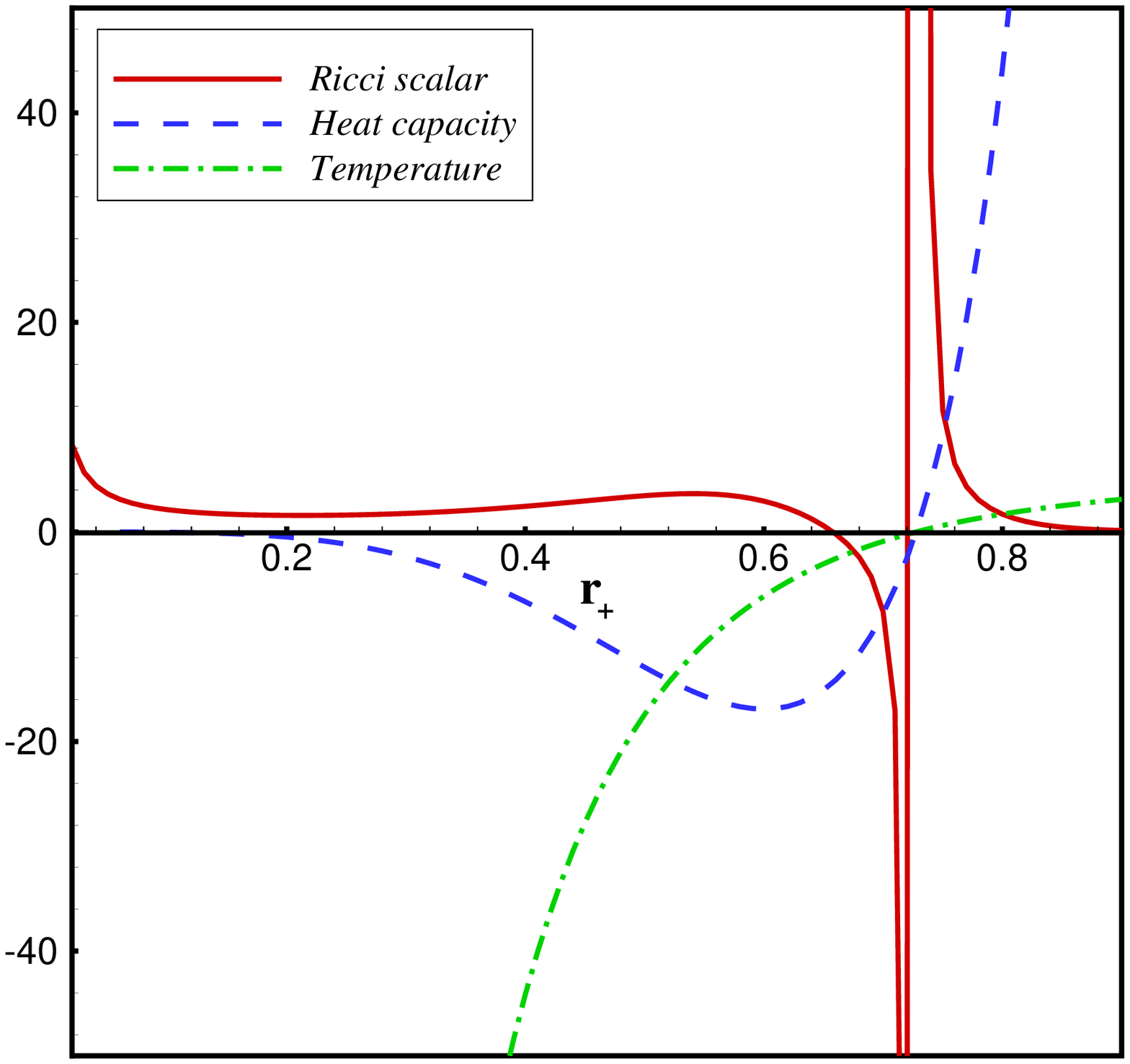}\label{fig6a}}\hspace*{.2cm}
\subfigure[$\Xi=0.9$]{\includegraphics[scale=0.27]{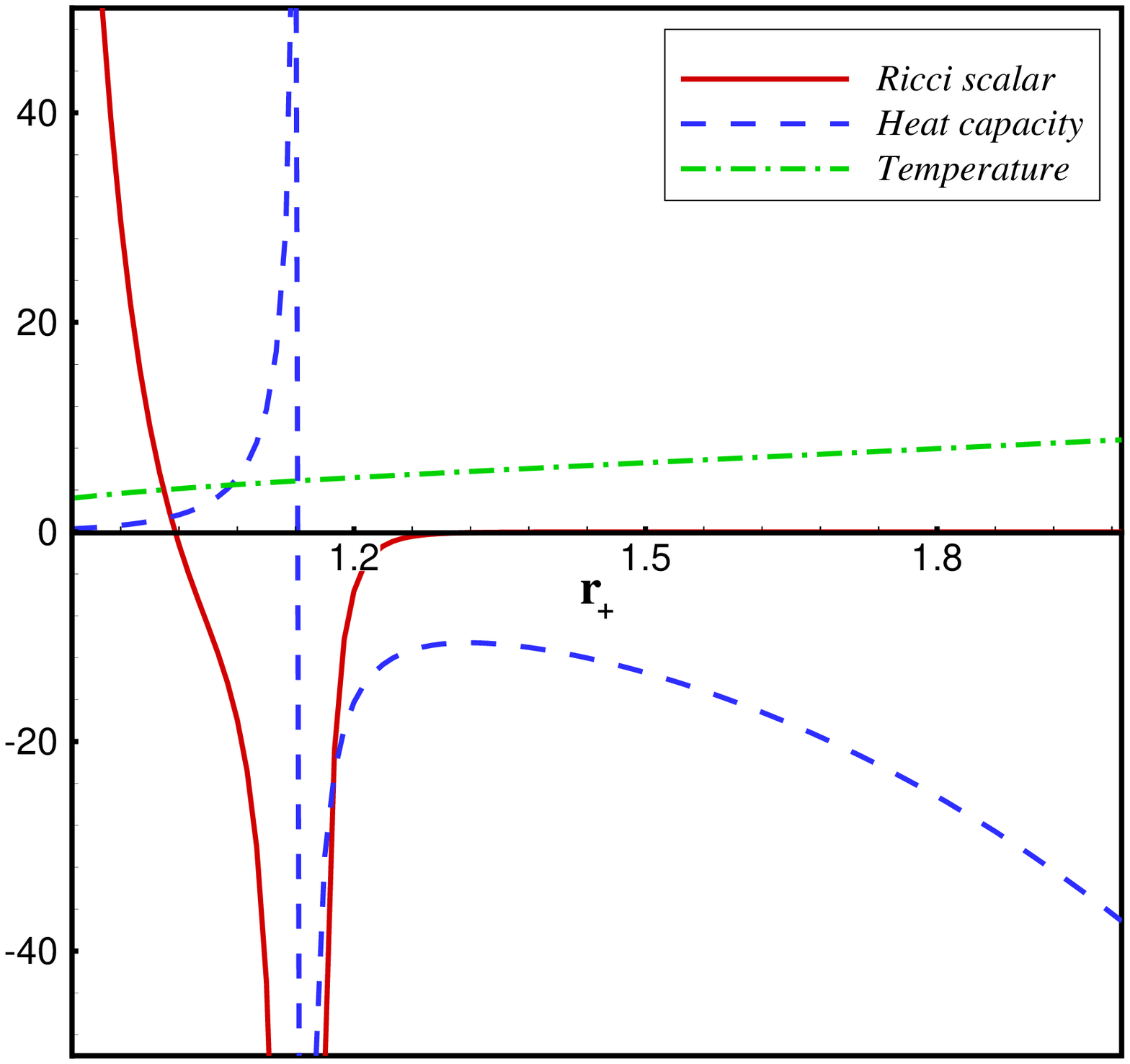}\label{fig6b}}\hspace*{.2cm}
\subfigure[$\Xi=1.1$]{\includegraphics[scale=0.27]{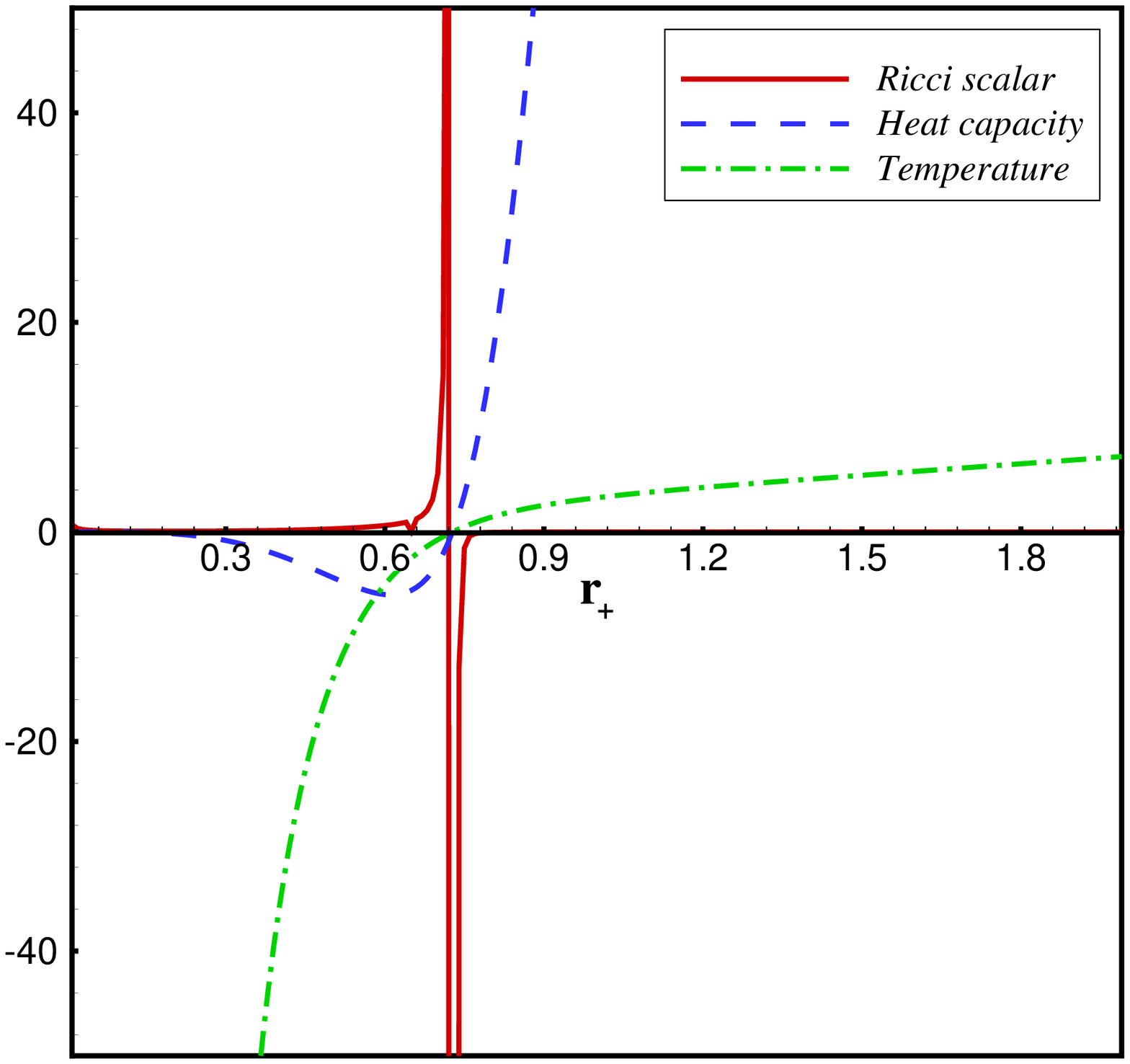}\label{fig6c}}\caption{Ricci scalar, heat capacity and temperature$T$ versus $r_{+}$ for AdS solutions with $q=1$, $\beta=2$ and $n=5$.}\label{figure6}
\end{figure}

\begin{figure}
\centering
\subfigure[$\Xi=0.9$]{\includegraphics[scale=0.27]{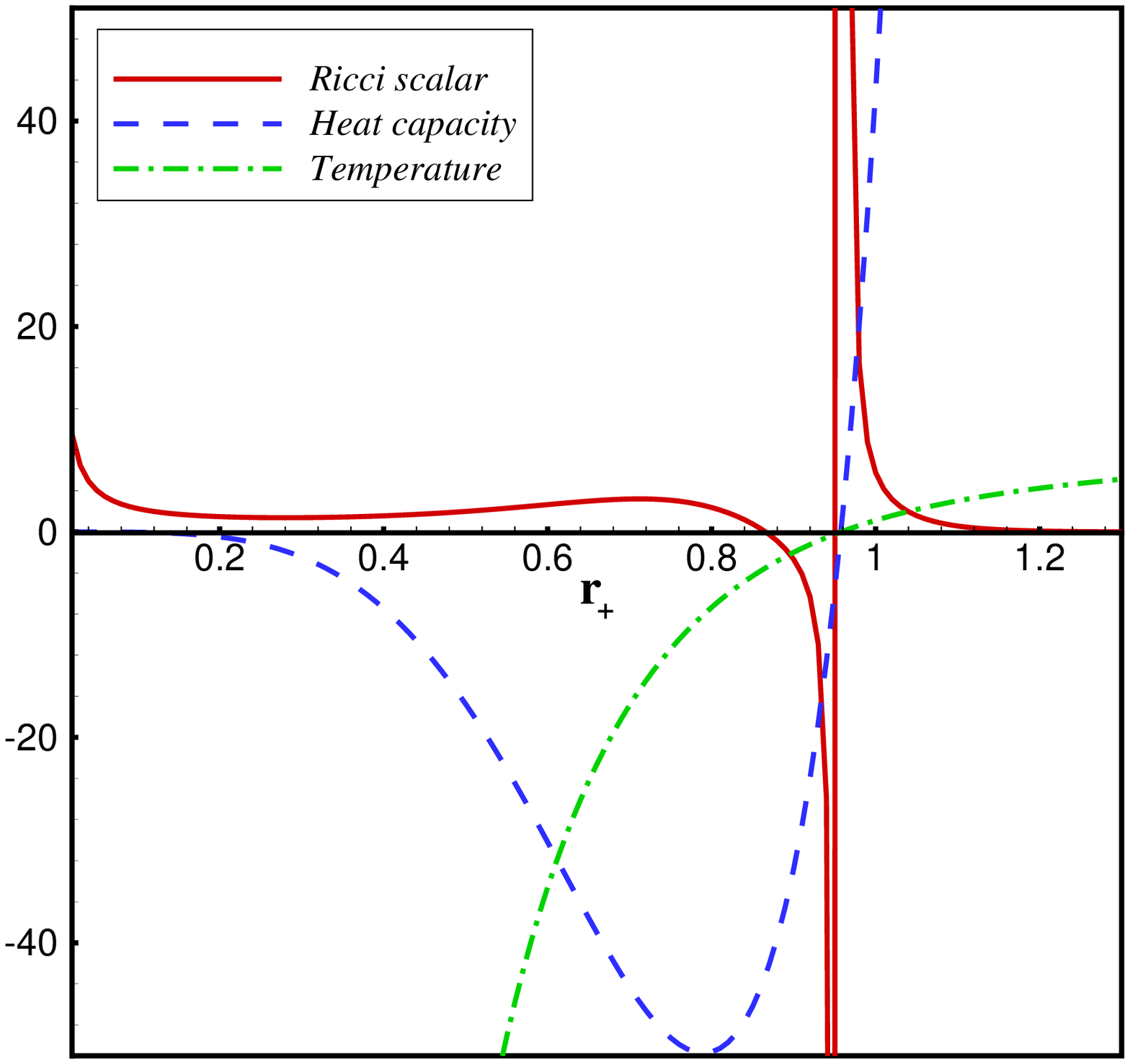}\label{fig7a}}\hspace*{.2cm}
\subfigure[$\Xi=0.9$]{\includegraphics[scale=0.27]{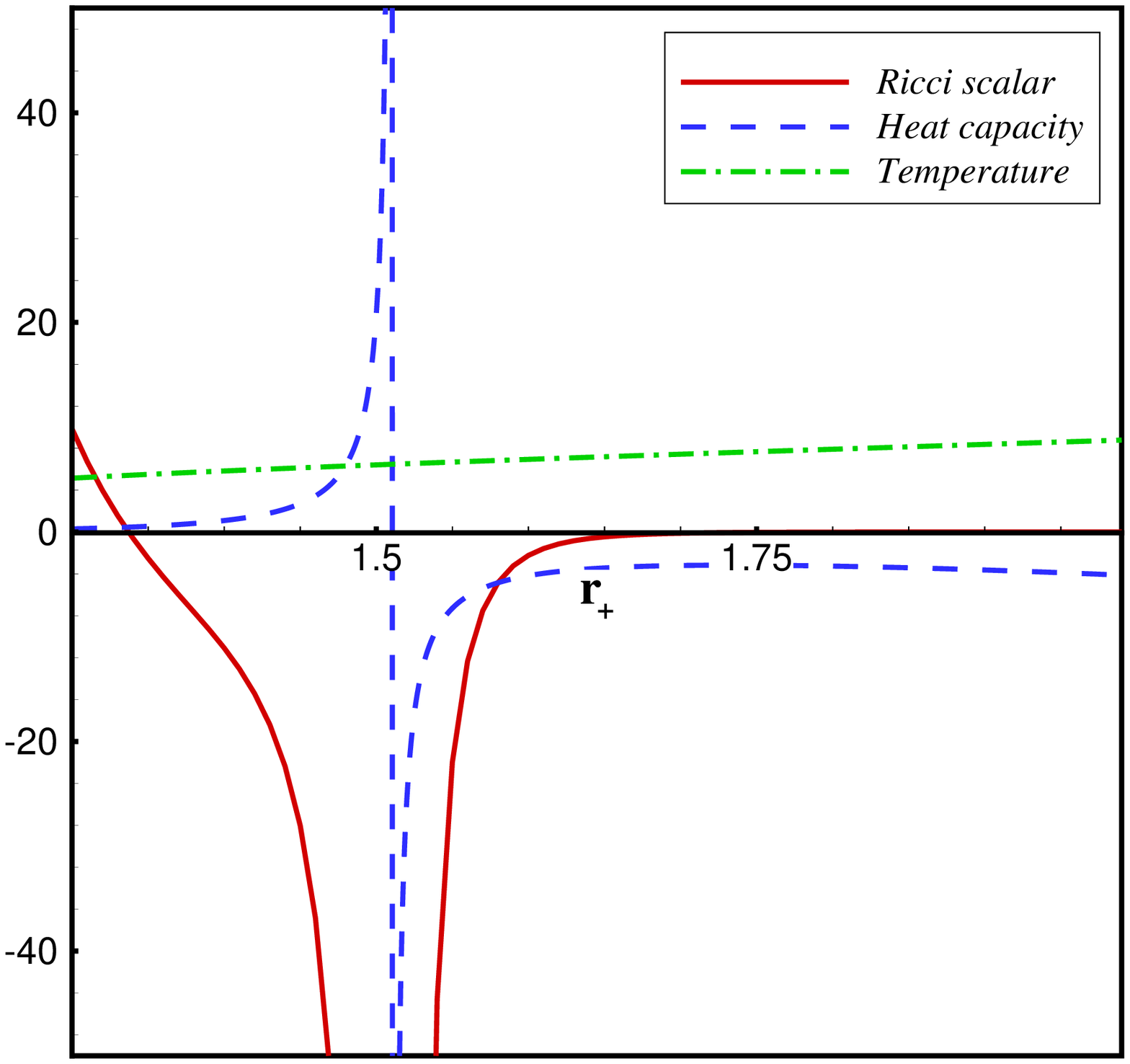}\label{fig7b}}\hspace*{.2cm}
\subfigure[$\Xi=1.1$]{\includegraphics[scale=0.27]{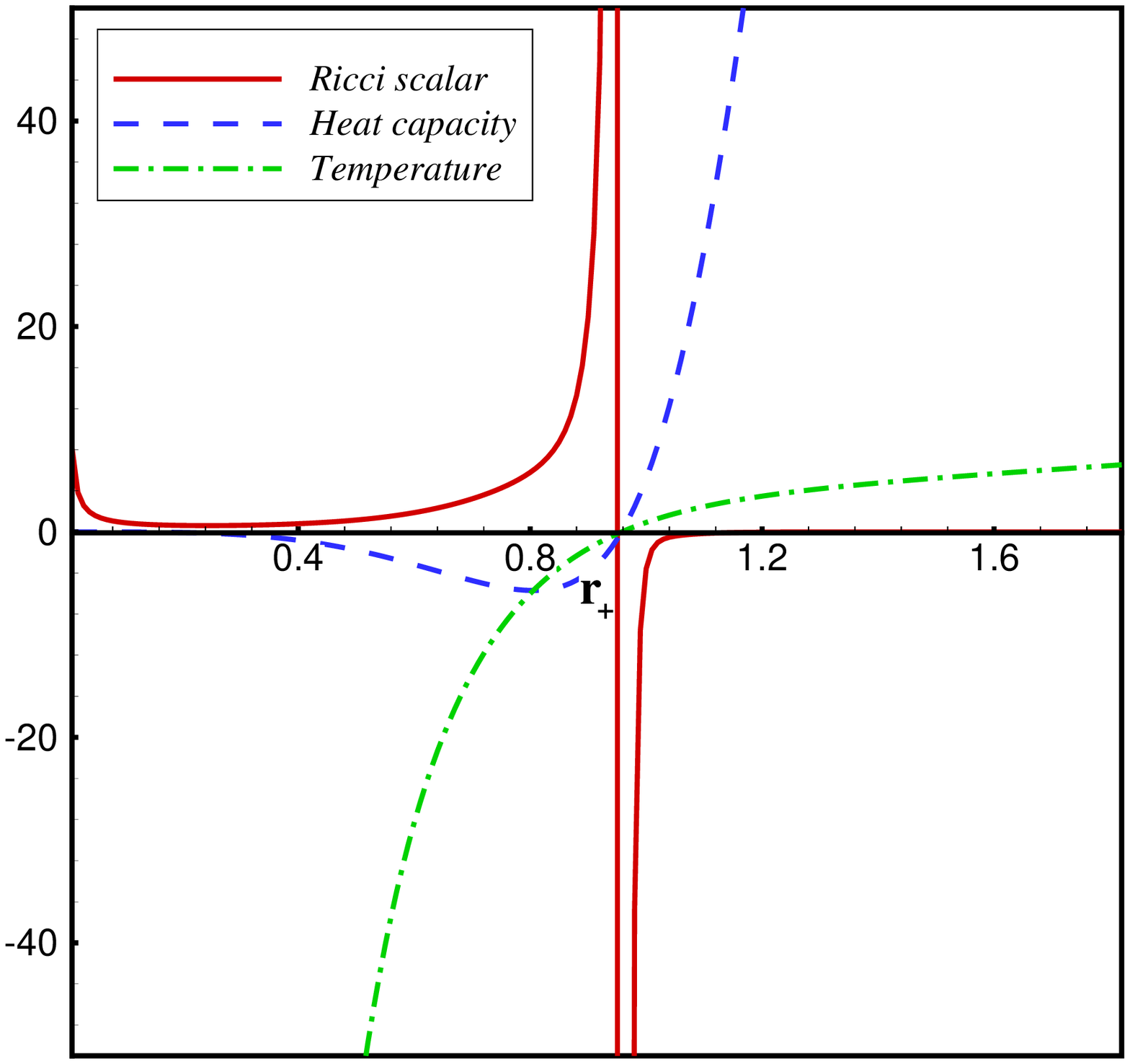}\label{fig7c}}\caption{Ricci scala, heat capacity and temperature$T$ versus $r_{+}$ for AdS solutions with $q=3$, $\beta=2$ and $n=5$.}\label{figure7}
\end{figure}

Now, we have the ability to extend our study to geometrothermodynamics or GTD. GTD is a method based on differential geometry concepts describing
properties of a thermodynamic system such as critical behavior and phase transition. Using a suitable metric as the geometry part, this formalism can act in a way that the infinite or zero values of the scalar curvature may match with the phase transition points of a thermodynamic system. The first approach for GTD was suggested separately by Weinhold \cite{Wein} and Ruppeiner \cite{Rup}. These two proposed metrics are conformally related to each other by the inverse temperature as the conformal factor. They have been also successful to describe the thermodynamical geometry of ordinary systems in \cite{Janyszek,Dolan} and to bring interesting results for black holes such as \cite{Cho,Aman}. Unfortunately, the thermodynamic results of these metrics are not invariant under the Legendre transformations and they depend on the choice of thermodynamic potentials \cite{Salamon}. After that, Quevedo proposed a new metric with Legendre invariant\cite{Que}. This method could not also
explain the correspondence between phase transitions and singularities of the scalar
curvature for some black holes \cite{Rodrigues}. This motivated Hendi et al. to introduce HPEM metric \cite{HPEM}
\begin{equation}\label{HPEM}
ds_{HPEM}^{2}=\frac{SM_{S}}{\left( \Pi _{i=2}^{n}\frac{\partial ^{2}M}{%
\partial \chi _{i}^{2}}\right) ^{3}}\left(
-M_{SS}dS^{2}+\sum_{i=2}^{n}\left( \frac{\partial ^{2}M}{\partial \chi_{i}^{2}}\right) d\chi _{i}^{2}\right) ,
\end{equation}
where $\chi _{i}$ ($\chi _{i}\neq S$) are extensive parameters and $M_{S} = \partial M/\partial S$, $M_{SS} = \partial ^2 M/\partial S^{2}$. The heat capacity is also defined as ${(\frac{M_{S}}{M_{SS}})}_{Q}$. Until now, this metric could have predicted the phase transition points correctly and the scalar curvature diverges exactly at the phase transition points in many black holes. Now, we are eager to use HPEM metric in this paper to see if it can predict the phase transition points correctly or not. For this purpose, we have plotted Ricci scalar of the metric \eqref{HPEM}, heat capacity and temperature of our solutions in Figs. \ref{figure6}-\ref{figure7}. We have refused to study GTD for dS and flat spacetimes, because as we said in the previous section, they are not physical. According to all of these figures, HPEM metric is successful to predict the divergence points of the Ricci scalar exactly at the phase transition points in which the heat capacity is zero or it diverges. There are two kinds of phase transition points which in the first one, the heat capacity is zero and the black brane has a transition from an unstable state (negative heat capacity) to a stable one (positive heat capacity). The temperature in this point is also zero. In the other kind, the heat capacity diverges and the temperature has a positive value. In this type, the black brane has a transition from a stable state to an unstable one. In Fig. \ref{figure6}, we have checked out GTD for diverse values of parameter $\Xi$. It is clear that for fixed parameters $q$, $\beta$, $n$ and $\Xi=0.9$, our solutions have two transitions which in the first one, the brane transits from an unstable state to a stable one and then, it has a transition to an unstable state in the second one. But, for larger values of parameter $\Xi$ ($\Xi=1.1$), the black brane has just one transition and moves to a stable state. In Fig. \ref{figure7}, we have repeated the behaviors of Fig. \ref{figure6} but for $q=3$. It shows that transitions and their behaviors are like the ones in Fig. \ref{figure6} but by increasing the value of $q$ in Fig. \ref{figure7}, the transitions happen in the larger $r_{+}$.
\section{concluding remarks}\label{con}
The idea of nonlinear electrodynamics was brought up for some reasons which the most important is the ability to remove the divergence of the electrical field at the origin. Nonlinear electrodynamics with quasitopological gravity is a new and interesting research which motivated us to investigate it. So, in this paper, we started our theory with quartic quasitopological gravity coupled to EN and LN forms of nonlinear electrodynamics. We obtained the solutions of this theory in two parts, static black hole solutions and rotating black brane ones. According to our expectation, for large values of nonlinear parameter $\beta$, the obtained solutions reduce to the solutions of quartic quasitopological gravity with linear Maxwell theory. Our solutions also have an essential singularity at $r=0$ and the function $f(r)$ goes to zero at this point. For fixed values of the parameters $n$, $\beta$, $m$ and $\hat{\mu}_{i=2,3,4}$, we can have an extreme black hole/brane for $q=q_{\rm{ext}}$ and a black hole/brane with two horizons for $q<q_{\rm{ext}}$ and a naked singularity for $q>q_{\rm{ext}}$. Therefore, the smaller values of $q$ can lead to a black hole with two horizons. Also, we concluded that the value of the horizons doesn't depend on the value of quasitopological parameters $\hat{\mu}_{i=2,3,4}$ and so, the horizons are independent of quasitopological gravity. They are related to the values of the parameters $q$, $\beta$, $n$ and $m$. For example, for fixed values of parameters $q$, $\beta$ and $n$, the value of $r_{+}$ in dS solutions increases, as the value of $m$ decreases. \\
Then, by using the Gibbons-Hawking method, we obtained the thermodynamic quantities of the solutions and by a Smarr-type formula, we proved that the solutions obey the first law of thermodynamics.
We also studied the thermal stability of the solutions in grand canonical ensemble. Unfortunately, dS and flat solutions are not physical because the temperature in these spacetimes is negative. But for AdS solutions, as the value of $det(H)$ is positive for all values of $r_{+}$, the thermal stability depends on the value of the temperature. There is a ${r_{+}}_{\rm{ext}}$ which the temperature is positive for $r_{+}>{r_{+}}_{\rm{ext}}$ and the value of ${r_{+}}_{\rm{ext}}$ is just dependent to the values of parameters $q$, $n$, $\beta$ and $m$. For example, by decreasing the value of $q$, the value of ${r_{+}}_{\rm{ext}}$ decreases. This manifests that for smaller values of parameter $q$, we have a larger region in which the temperature is positive and thermal stability is established. \\
We also studied GTD for the solutions of quartic quasitopological gravity with nonlinear electrodynamics. We used HPEM metric and demonstrated that it has the ability to predict the divergences of the Ricci scalar exactly at the phase transition points. We found two kinds of transitions which in the first type, the heat capacity and temperature are both zero and in the second one, the heat capacity diverges while the temperature has a finite value. For small values of parameter $\Xi$, the black brane has two transition points for fixed values of other parameters and it finally transits to an unstable state. But for larger $\Xi$, there is just one transition which takes the brane to a stable state. Also, by increasing the value of the parameter $q$, the transitions happen in larger $r_{+}$. \\
It should be noted that we can extend this study to quintic quasitopological gravity with or without nonlinear electrodynamics. We can also extend our study to a theory of quartic-quasitopological gravity coupled to nonlinear electrodynamics in Lifshitz spacetime.
\section{Appendix}
\subsection{coefficients of quartic quasitopological terms}\label{app}
The $c_{i}$'s for ${{\mathcal L}_4}$ in Eq. \eqref{quasi4} are defined as
\begin{eqnarray}
&&c_{1}=-(n-1)(n^7-3n^6-29n^5+170n^4-349n^3+348n^2-180n+36)\nonumber\\
&&c_{2}=-4(n-3)(2n^6-20n^5+65n^4-81n^3+13n^2+45n-18)\nonumber\\
&&c_{3}=-64(n-1)(3n^2-8n+3)(n^2-3n+3)\nonumber\\
&&c_{4}=-(n^8-6n^7+12n^6-22n^5+114n^4-345n^3+468n^2-270n+54)\nonumber\\
&&c_{5}=16(n-1)(10n^4-51n^3+93n^2-72n+18)\nonumber\\
&&c_{6}=-32(n-1)^2(n-3)^2(3n^2-8n+3)\nonumber\\
&&c_{7}=64(n-2)(n-1)^2(4n^3-18n^2+27n-9)\nonumber\\
&&c_{8}=-96(n-1)(n-2)(2n^4-7n^3+4n^2+6n-3)\nonumber\\
&&c_{9}=16(n-1)^3(2n^4-26n^3+93n^2-117n+36)\nonumber\\
&&c_{10}=n^5-31n^4+168n^3-360n^2+330n-90\nonumber\\
&&c_{11}=2(6n^6-67n^5+311n^4-742n^3+936n^2-576n+126)\nonumber\\
&&c_{12}=8(7n^5-47n^4+121n^3-141n^2+63n-9)\nonumber\\
&&c_{13}=16n(n-1)(n-2)(n-3)(3n^2-8n+3)\nonumber\\
&&c_{14}=8(n-1)(n^7-4n^6-15n^5+122n^4-287n^3+297n^2-126n+18).\nonumber\\
\end{eqnarray}
\subsection{details of quartic quasitopological black hole solutions}\label{app1}
\begin{eqnarray}
P&=&-\frac{\alpha^2}{12}-\gamma\,\,\,\,\,\,\,\,\,\,,\,\,\,\,\,\,\,
H=-\frac{\alpha^3}{108}+\frac{\alpha \gamma}{3}-\frac{\beta^2}{8},
\end{eqnarray}
that $\alpha$, $\beta$ and $\gamma$ are
\begin{eqnarray}\label{eq1}
\alpha&=&\frac{-3\mu^2}{8c^2}+\frac{\lambda}{c}\,\,\,\,\,\,\,\,\,\,\,,\,\,\,\,\,\,
\beta=\frac{\mu^3}{8c^3}-\frac{\mu\lambda}{2c^2}+\frac{1}{c}\nonumber\\
&&\gamma=\frac{-3\mu^4}{256c^4}+\frac{\lambda \mu^2}{16c^3}-\frac{\mu}{4c^2}+\frac{\kappa}{c}.
\end{eqnarray}

If we define the following definitions,
\begin{eqnarray}
U=\bigg(-\frac{H}{2}\pm\sqrt{\Delta}\bigg)^{\frac{1}{3}},
\end{eqnarray}
\begin{equation}
y=\left\{
\begin{array}{ll}
$$-\frac{5}{6}\alpha+U-\frac{P}{3U}$$,\quad \quad\quad\quad \  \ {U\neq 0,}\quad &  \\ \\
$$-\frac{5}{6}\alpha+U-\sqrt[3]{H}$$, \quad \quad\quad\quad{U=0,}\quad &
\end{array}
\right.
\end{equation}
\begin{eqnarray}
W=\sqrt{\alpha+2y},
\end{eqnarray}

\acknowledgments{We would like to thank Payame Noor University and Jahrom
University.}
%%%%%%%%%%%%%%%%%%%%%%%%%%%%%%%%%%%%%%%%%%%%%%%%%%%%%%%%%%%%%%%%%%

\end{document}